\documentclass{article}

\usepackage{caption}
\usepackage{amsmath}
\usepackage{amssymb}
\usepackage{amsfonts}
\usepackage{algorithmic}
\usepackage{textcomp}
\usepackage{xcolor}

\usepackage{enumitem}
\usepackage{xspace}
\usepackage{graphicx}
\usepackage{comment}

\graphicspath{{./figures/}}

\usepackage[dvipsnames]{xcolor}
\usepackage{booktabs}

\usepackage{hyperref}
\hypersetup{
    unicode=true,
    bookmarksnumbered=true,
    colorlinks=true,
    linkcolor={red!70!black},
    citecolor={red!70!black},
    urlcolor={blue!70!black},
    pdfborder={0 0 0}
}
\usepackage[capitalize,nameinlink]{cleveref}

\newcommand{\name}{ConServe}
\newcommand{\systemname}{ConServe}

\newcounter{insight}

\newcommand{\newedit}[1]{#1}
\newcommand{\todo}[1]{\textcolor{red}{\textbf{TODO:} #1}}

\newcommand{\myparagraph}[1]{\vspace{\smallskipamount}\noindent\textbf{#1.\xspace}}

\newcommand{\ie}{\emph{i.e.}\xspace}

\newcommand*{\rom}[1]{\uppercase\expandafter{\romannumeral #1\relax}}

\definecolor{DarkRed}{HTML}{8B0000}       
\definecolor{ForestGreen}{HTML}{228B22}   

\usepackage{multiaudience}

\newif\ifshowcomment
\showcommenttrue

\ifshowcomment
    \SetNewAudience{comments}
    \newcommand{\jerry}[1]{{\color{teal}[JD: #1]}} 
\else
    \SetNewAudience{}
    \newcommand{\jerry}[1]{\ignorespaces}
    \newcommand{\todo}[1]{\ignorespaces}
\fi

\DefCurrentAudience{comments}

\newcommand{\mingyuan}[1]{\showto{comments}{\textcolor{red}{{\textbf{Mingyuan}:} \textsl{#1}}}}
\newcommand{\pouya}[1]{\showto{comments}{\textcolor{purple}{{\textbf{pouya}:} \textsl{#1}}}}

\usepackage[accepted]{mlsys/mlsys2026}
\mlsystitlerunning{Let the computation be where it should be}

\begin{document}
\twocolumn[
\mlsystitle{Observation, Not Prediction: Conversation-Level Disaggregated Scheduling for Agentic Serving}

\mlsyssetsymbol{equal}{*}

\begin{mlsysauthorlist}
\mlsysauthor{Jianru Ding}{inst1}
\mlsysauthor{Ryien Hosseini}{equal,inst1}
\mlsysauthor{Pouya Mahdi Gholami}{equal,inst1}
\mlsysauthor{Mingyuan Xiang}{equal,inst1}
\mlsysauthor{Henry Hoffmann}{inst1}
\end{mlsysauthorlist}

\mlsysaffiliation{inst1}{University of Chicago} 
\mlsyscorrespondingauthor{Jianru Ding}{jrding@uchicago.edu}

\vskip 0.3in
\begin{abstract}
\comment{
LLM-based agents complete complex tasks by interleaving model inference with tool calls across many turns, producing a workload with large initial context, growing KV state reused across turns, and latency measured at job completion rather than per request. These characteristics expose two failure modes: collocated prefill-decode engines suffer interference as incremental per-turn prefills compete with ongoing decoding; \mingyuan{This part contradicts with later claim. The real problem is that "collocated" is misused when prefill has high costs. I would repharse to "blindly collocated prefill-decode engines suffer interference when prefills compete with ongoing decoding, e.g, for the first-turn prefill"} full PD disaggregation accumulates KV transfer overhead across every turn. Meanwhile, GPU power demands are growing faster than energy infrastructure can support, making heterogeneous deployments — pairing high-power GPUs with their lower-power predecessors — an increasingly practical necessity. \mingyuan{The power wall of data center doesn't necessarily lead to heterogeneous GPUs, for example, power gating or better cooling can also reduce power.} 
\name{} exploits this heterogeneity by disaggregating only the first-turn prefill to a high-power GPU, then pinning all subsequent computation to lower-power decoders with cached KV state. We show that this adaptive policy achieves X\% reduction in end-to-end job completion latency and Y\% improvement in energy efficiency, and characterize the load-dependent crossover at which it outperforms both collocated and fully disaggregated baselines.
}

\comment{
\newedit{
LLM-based agents complete complex tasks across many turns of model inference and tool calls, producing workloads whose total latency is fundamentally unknowable when the job arrives. Scheduling such workloads efficiently therefore requires either predicting duration or designing systems that do not need predictions. Examples of the latter include systems that make per turn decisions about routing requests to prefill or decode nodes. These designs share a common limitation: ther treating each turn as an independent routing decision, optimizing local, per-turn metrics that may not reflect the aggregate job completion time. Our key insight is that the right unit of scheduling is the conversation, not the turn: by disaggregating only the turn-1 prefill and pinning each conversation to a decoder replica for its lifetime, turn-1 input length becomes the sole signal needed for admission control and load balancing, and it is fully observable at arrival. We implement this insight in \name{}, which routes turn-1 prefill to a high-throughput prefill node  provisioned as the system bottleneck, transfers the resulting KV state once to a decoder replica, and pins each conversation to that replica for all subsequent turns with full cache reuse. The design maps naturally onto heterogeneous clusters that pair high-power prefillers with memory-bandwidth optimized decoders. \name{} reduces end-to-end job completion latency by 28.0\% and improves energy efficiency by 7.0\% over both per-turn-routing and fully disaggregated baselines, and we characterize the load-dependent crossover regime in which each design wins.
}
}

LLM-based agents resolve a user task through many turns of dependent inference and tool calls, producing a workload whose total cost is unknown when the task arrives. Existing multi-turn systems keep the turn as the scheduling unit and decide, turn by turn, whether to disaggregate prefill from decode. That decision rests on the turn's decode length, tool behavior, and KV growth, quantities that are not observable when the scheduler must act, forcing the system to predict them. We show this dependence on prediction is imposed by the scheduling unit, not the workload.
Raising the scheduling unit from the turn to the conversation converts turn-level irregularity into a stable, two-phase structure: 1) a compute-bound turn-1 prefill followed by 2) a long, memory-bound tail. Thus, with the conversation as the scheduling unit, placement reduces to reading the first-turn input length and per-decoder KV occupancy, both directly observable.  We instantiate this principle in ConServe, which routes the first-turn prefill to a high-throughput prefiller, transfers the KV cache exactly once, and pins the conversation to a single decoder for its entire tail, with no learned model of decode-side cost. Against a per-turn prediction baseline, ConServe reduces p95 time-to-first-effective-token (the latency of a conversation's first user-visible output) by 51.08\% and improves energy efficiency by 7.51\% while preserving last-turn TBT and SLOs; mapping the two phases onto heterogeneous GPU tiers adds a further 22.75\% in energy efficiency.

\end{abstract}
]

\printAffiliationsAndNotice{\mlsysEqualContribution}


\thispagestyle{plain}
\pagestyle{plain}

\section{Introduction}
\label{sec:intro}

Traditional large language model (LLM) serving is built around a simple contract: a request arrives, the model prefills its prompt and decodes a response, and the request departs \cite{transformer}. Agentic workloads break this contract. A user task is no longer resolved by a single model invocation but by a sequence of dependent inference calls interleaved with tool use, environmental feedback, and accumulating state \cite{ReAct}. The served object is therefore not a request but a \textit{stateful, multi-turn program} which moreover exhibits a characteristic temporal asymmetry: its first turn encodes a long task description (e.g., instructions, repository state, retrieved context) in a single compute-bound prefill, while later turns typically append only short tool outputs and observations, dominated by memory-bound decoding over a key-value (KV) cache that only grows. A conversation is therefore, typically, structured as one heavy prefill followed by a long, light, memory-bound tail.

This shift is not merely quantitative. Current generic LLM serving systems typically take the request (equivalently, a single prefill--decode turn) as the atomic unit of scheduling. For single-shot inference this approach is harmless, as the unit the system schedules coincides with the unit the user values. Agentic workloads break this coincidence. What the user values is the conversation's final, externally meaningful output, yet most intermediate turns emit only tool calls that are never read by the user. The result is a \textit{scheduling--value mismatch}: the system optimizes per-turn latency while the user waits on a conversation-level response. Consequently, existing serving systems that optimize each turn in isolation optimize the wrong objective.
To name the value the user actually cares about, we introduce \textbf{time-to-first-effective-token} (TTFET): the time until the first token of genuine, user-visible progress, as opposed to the internal tokens that drive tool calls. TTFET is thus a property of a conversation, not of any turn.

Existing multi-turn serving systems do recognize that later turns differ from the first. Building on prefill--decode disaggregation, which separates the compute-bound prefill from latency-sensitive decoding onto distinct workers \cite{DistServe}, these systems add append-prefill routing, KV-cache retention, and tool-aware scheduling to avoid recomputing shared state across turns \cite{continuum, infercept, autellix}. Yet these designs share a single design principle: the decision unit remains the turn. For each turn, such a system decides whether to disaggregate it (i.e., whether to pay a remote prefill and a KV transfer or process it locally) and reaches that decision by profiling or predicting the turn's cost \cite{ampd, ppd}.

In this work, we first argue that this per-turn scheduling decision is \textit{structurally brittle}. The cost that drives the decision cannot be known in advance: a turn's prefill is fixed by its input length and can be read directly, but other key behavior (e.g., how many tokens it will decode, whether it will call a tool, and how much it will enlarge the KV cache) is unobservable at the moment the scheduler must act. Yet such decode-side quantities are what determine whether disaggregating the turn pays off, so per-turn placement must predict them. Any predictor, however well-tuned, will eventually misjudge a turn and route it incorrectly (we characterize this prefill/decode predictability gap empirically in Section~\ref{sec:characterization}). We argue this prediction is not a requirement of the workload but of the scheduling unit: partitioning a conversation into independently scheduled turns discards precisely the structure that would have made placement deterministic. \emph{Thus, in prior work, the unit, not the predictor, is the source of the fragility.}

\myparagraph{Key Insight}
We schedule the conversation, not the turn. At turn granularity, agentic serving is irregular: decode lengths, tool calls, append-prefill sizes, and KV growth are difficult to infer when placement decisions are made. At conversation granularity, \textit{this irregularity collapses into a stable two-phase structure}: an initial compute-bound prefill that materializes task state, followed by a memory-bound trajectory consisting of the first decode and all subsequent append-prefill/decode turns. Later turns may contain append-prefill work, but their input is small relative to the accumulated KV cache, so execution is dominated by KV movement and attention over existing state. 
This restores the original prefill--decode abstraction but at conversation-level granularity: one compute-bound phase, one memory-bound phase, and a KV transfer at the boundary. 

Two consequences follow. First, scheduling need not predict turn-level outcomes such as future decode lengths or tool behavior; it can instead rely on the same kind of observable phase structure that classical serving systems already expose. Second, heterogeneous serving follows by construction: under heterogeneous cluster environments, the same phase split maps compute-intensive initial prefills to high-throughput GPUs and the long memory-bound tail to devices with sufficient memory bandwidth and capacity.


\myparagraph{\systemname{}}
We instantiate this principle in \systemname{}\footnote{So named as to reflect two commitments: ConServe \textit{serves} stateful, multi-turn agentic \textit{con}versations and \textit{conserves} energy by spending it only on the phase that requires it.}, a conversation-level disaggregation scheduler for agentic serving. \systemname{} mirrors the original prefill--decode abstraction at the conversation level: it routes the first-turn prefill as the compute-bound phase, transfers the resulting KV cache exactly once, and then runs the first-turn decode and every subsequent append-prefill/decode turn as a single memory-bound phase. Under \systemname{}, a conversation is therefore pinned for its entire memory-bound tail, avoiding repeated KV movement across turns.
This paradigm shows that \textit{placement should be reactive rather than predictive.} New conversations are assigned using observable quantities: input-token counts for the initial prefill and decoder KV-cache utilization for the memory-bound tail. \systemname{} overprovisions decode replicas to absorb variation in conversation length; when a prefiller is saturated, or when decoders approach saturation under bursts of long conversations, the scheduler routes the \emph{next} conversation elsewhere rather than migrating one already placed.

We evaluate \systemname{} on representative agentic workloads \cite{swebench}, using a  Qwen3-0.6B model backbone and a NVIDIA A40 GPU cluster, against collocated, fully disaggregated, and per-turn prediction-based baselines. Compared to the per-turn baseline \cite{ampd}, \systemname{} reduces p95 TTFET by 51.08\% and improves energy efficiency by 7.51\% while maintaining similar last-turn time-between-tokens (TBT) and preserving service-level objectives (SLOs). It does so with a deliberately minimal policy: one disaggregation per conversation and no learned cost model. On heterogeneous hardware, \systemname{} preserves these latency results while improving energy efficiency by a further 22.75\% over its homogeneous configuration. These gains show that the relevant abstraction is the conversation, not the turn. We therefore present conversation-level scheduling as a principled paradigm for agentic serving, and \systemname{} as a minimal instantiation of this principle. 
We make the following contributions:
\begin{itemize} [leftmargin=1em]
\item \textbf{Conversation-level scheduling.} We identify a \emph{scheduling--value mismatch} in agentic serving: systems optimize per-turn latency, while users wait for conversation-level outcomes. We argue that the conversation is therefore the correct unit of scheduling, and introduce \emph{time-to-first-effective-token} (TTFET) and \emph{last-turn time-between-tokens} (TBT) as metrics that capture this objective. Raising the unit to the conversation replaces turn-level prediction with directly observable state, eliminating the source of brittleness in prior multi-turn schedulers.
%
%
\item \textbf{\systemname{}.} We introduce a disaggregation scheduler that routes first-turn prefills to high-power GPUs, transfers KV state exactly once, and executes a conversation's entire tail locally on a separate (possibly low-power) decoder. Committing to one placement per conversation from observable signals allows 
\systemname{} to work without a learned cost model.
\item \textbf{Empirical demonstration of structural brittleness.} We show that per-turn scheduling is brittle in a structural sense: under a per-turn prediction baseline, SLO violations and energy efficiency degrade linearly in prediction error rate, while ConServe's conversation-level placement holds constant by construction.  
%
\item \textbf{Heterogeneity and energy as corollaries.} Because placement reads only observable states, tier-aware allocation and conversation-level energy optimization follow without additional machinery, and both are largely unaddressed by prior multi-turn systems.
%
\end{itemize}

\section{Background}

\comment{
\pouya{@Mingyuan, prior work needs to cover the following narrative. Some of these might already be covered, but I'm writing this holistically. Lmk what you think and see how it matches with my narrative in section 3. I'm going to make section 3 mostly about the factors that impact the important metrics.}

{\color{red}

\begin{itemize}
    \item LLM inference: What it is and how do we quantify the goodness of its serving (i.e. metrics such as ttft and tpt).
    Next discuss factors that impact these metrics, i.e. KV cache, Prefill, Decode, Prefix Cache (I think the factors are covered, but no the metrics. Might need to turns this on its head to make it work.). 
    
    \item Agentic inference: builds on LLM inference, but has a different aim. discuss what the different aim changes in how the operation works: the multi turn setup, large initial input, small appendages per turn, long context length. The aim and the structure of the operation change the traditional metric to the jerry metric. Then forward reference to section 3 discussing the factors that impact the jerry metric. INTRODUCE THE CONEPT OF CONVERSATION?
    
    \item Prior work on LLM and Agentic Serving: How LLM and Agentic serving are done. We need to hammer that prior agentic servings are dumb. Full PD disagg is stupid cause no cache (forward reference); collocaiton is dumb cause first prefill bad; mix methods are profiling based and well, that's even more dumb. Cherry on top they all use the wrong metrics; you gotta rig it up to use the jerry metrics, those are the good ones.
    
    \item Serving on Heterogeneous GPUs: Hold on this until I make sure we're clear on the messaging.
\end{itemize}
}
}

\comment{
\subsection{LLM Inference and Serving}

Transformer-based language models \cite{transformer} generate text autoregressively, producing one token per forward pass until a stopping condition is met \cite{gpt-1, gpt-2}. The first forward pass---the prefill stage---processes all input tokens in parallel, computing a key-value (KV) cache of intermediate attention states for every input token. The decode stage then generates new tokens one at a time, reusing the KV cache from all prior tokens and appending a new entry at each step. Efficient management of this growing KV cache is critical for serving throughput \cite{vLLM}. Prefill is consequently compute-bound, processing many tokens simultaneously; decode is memory-bound, performing minimal computation per step but loading the entire growing KV cache from memory at every forward pass \cite{PopeDCDBHXAD23, Splitwise}. Prefix caching further exploits this structure: when a new request shares a common prefix with a previously cached request, the prefill computation for those tokens is skipped entirely, directly reducing both latency and compute cost \cite{SGLang, CacheGen, CacheBlend}.

LLM serving quality is often evaluated on \textbf{Time-to-first-token (TTFT)} and \textbf{Time-between-tokens (TBT)}. TTFT measures the delay from request arrival to the emission of the first output token, \emph{reflecting how long a user waits before seeing any response}. Time-between-tokens (TBT) measures the inter-token delay during generation. Existing serving systems try to minimize these two metrics without violating service-level objectives (SLOs).

Prior work has studied extensively how to improve LLM serving throughput. Iteration-level scheduling \cite{Orca} batches forward passes from different requests at each step, allowing finished requests to leave and new ones to join the batch continuously. However, mixing prefill and decode in the same batch introduces contention: the compute-heavy prefill disrupts memory-bound decode, inflating time-between-tokens \cite{Sarathi-Serve, DistServe, VIDUR}. Chunked prefill \cite{Sarathi-Serve, Deepspeed-fastgen} mitigates this by splitting long prefills into smaller chunks interleaved with decode iterations, reducing stalls but not fully eliminating contention. Prefill-decode (PD) disaggregation \cite{DistServe, Splitwise, Mooncake} eliminates contention entirely by assigning prefill and decode to separate model replicas, at the cost of transferring KV states over the network between phases. In practice, collocated deployments combine iteration-level scheduling with chunked prefill to balance throughput and SLO compliance on a single set of replicas \cite{vLLM}, while disaggregated deployments eliminate interference at the cost of KV transfer overhead.
}

\newedit{
\subsection{LLM Serving Techniques}
Transformer-based language models \cite{transformer} generate text autoregressively in two stages \cite{gpt-1, gpt-2}. Prefill processes the input prompt in a single forward pass and produces a key-value (KV) cache for every input token. Decode then generates each subsequent token by reading the full KV cache and appending one new entry. Prefill is consequently compute-bound; decode is memory-bound, gated by KV cache bandwidth at every step \cite{PopeDCDBHXAD23, Splitwise}. Efficient management of this growing KV cache is critical for serving throughput \cite{cachedattention}.

A few techniques have been proposed and adopted in practice to address these characteristics and improve serving quality for user-facing LLM inference requests. Prefix caching \cite{SGLang} reuses the KV cache across requests sharing a common input prefix, eliminating redundant prefill computation. Dynamic batching \cite{orca} batches forward passes from different requests at each iteration so finished requests can leave and new requests can join continuously, raising GPU utilization. Its downside is that a long prefill batched into ongoing decode iterations stalls the decoders for the prefill's duration; chunked prefill \cite{Sarathi-Serve} mitigates this by splitting long prefills into smaller chunks that interleave with decode, bounding the per-step stall. Prefill-decode (PD) disaggregation \cite{DistServe, Splitwise} eliminates the problem entirely by assigning prefill and decode to separate replicas, exploiting the compute-bound vs memory-bound asymmetry to raise aggregate GPU utilization at the cost of transferring KV state between replicas.
}

\subsection{Multi-turn Agentic Conversations}
\label{sec:bg-agentic}
Agentic LLM applications \cite{LangChain, AutoGPT} follow a multi-turn ReAct execution loop, in which the model reasons about a user request prompt, emits a structured tool call (e.g., an API invocation, code execution, or database query), suspends until the tool returns, and resumes generation with the tool result appended to the context \cite{ReAct, Toolformer, Reflexion, ToolLLM}. We refer to the full sequence of turns required to complete a single user request as a \emph{conversation}. A conversation begins with a detailed, initial prompt that establishes the task and environment context, followed by a variable number of turns, each consisting of an LLM generation step and a subsequent tool invocation whose result is fed back as additional input \cite{AgentBench}. 
\newedit{
To improve context efficiency, modern agentic systems frequently restrict tool-call outputs to task-relevant information, avoiding the inclusion of complete files, logs, or other large artifacts in the prompt~\cite{anthropic_claude_code_limits_2026}.
}

\comment{
Subsequent turns only append short tool-call results to the existing context. Because the multi-turn execution is invisible to the user and only the final turn's output matters, traditional per-request metrics are misaligned. TTFT measured at the first turn captures latency for an intermediate tool call, not the response the user actually sees. Similarly, per-request TBT aggregates across all turns indiscriminately, when only the last turn's token delivery and the overall \textbf{end-to-end (E2E)} completion time are meaningful. We address this evaluation misalignment in Section~\ref{sec:characterization}. \mingyuan{update to specific section when section 3 is done} \pouya{I'm not sure whether we're addressing this in section 3 or just saying we use something else. TBD.}
}

Unfortunately, neither collocated nor fully disaggregated
prefill-decode architectures are well matched to this workload structure. In a collocated deployment, chunked prefill \cite{Sarathi-Serve} splits the long prefill into smaller chunks interleaved with decode steps, but a turn-1 prompt spanning thousands of tokens still occupies the GPU for many chunk iterations, delaying collocated decodes throughout the chunking window. Full prefill-decode disaggregation eliminates this interference by routing prefill and decode to dedicated GPUs \cite{DistServe, Splitwise}. In a multi-turn setting, however, this design routes every turn through the prefill node, disregarding a key structural property of agentic workloads: turn 2+ appends are short incremental prefills that could run directly on the decoder holding the cached prefix, with minimal contention (Figure \ref{fig:collocation}). Full disaggregation forfeits this optimization opportunity in turn 2+ requests.

\newedit{
Two concurrent systems route incremental prefills in multi-turn disaggregated serving, and both retain the turn as the decision unit. AMPD \cite{ampd} chooses, for each Turn 2+ prefill, between local execution and remote execution on a prefill node using real-time queue states and an offline cost model; remote execution reads the history KV from the decoder and writes the new KV back, a bidirectional transfer that departs from the one-way producer-consumer contract of standard disaggregation. PPD~\cite{ppd} makes the same per-turn choice through an offline lookup table indexed by context length, input-output ratio, and load. Keeping the turn as the unit has two consequences. First, each decision conditions on quantities unavailable when it is made: AMPD's cost model omits the collocated-decode interference that dominates local execution cost, and PPD's table is indexed by a request's output length, unknown at arrival. Second, both optimize per-turn TTFT and TBT, and neither measures TTFET or last-turn TBT, the conversation-level metrics that determine user-perceived performance (§\ref{sec:intro}). Both consequences trace to scheduling the turn rather than the conversation. The case for per-turn routing is also narrowest where agentic workloads sit: PPD's dynamic routing is motivated by Turn 2+ requests with heavy prefills, but agentic appends are uniformly short (§\ref{sec:characterization}), the regime in which local execution already dominates and the per-turn decision collapses to a fixed local policy. \systemname{} adopts that policy at conversation granularity, conditioning only on observable state and optimizing the metrics users experience.
}

\subsection{Serving with Heterogeneous GPUs}

\comment{
Several recent systems address LLM serving on heterogeneous GPU clusters, targeting general, single-turn workloads. These systems formulate GPU allocation as constrained optimization over cost or throughput. They either target colocated deployments, where the objective is to assign the right GPU type to each model replica or partition \cite{Melange, Helix, HexGen}, or disaggregated deployments, where prefill and decode phases are routed to separate hardware tiers \cite{HexGen-2, ThunderServe}. Although these approaches differ in optimization formulation and model partitioning strategy, they share a common insight: LLM inference places non-uniform demands on compute, memory capacity, and memory bandwidth, making it profitable to map workload to GPU capabilities rather than deploying a single homogeneous GPU type. However, all of these systems assume stateless, single-turn requests and therefore make routing decisions independently per request, with no consideration of cross-turn KV cache reuse or end-to-end job completion time.

Prior work has also explored serving optimizations for multi-turn and agentic workloads. Parrot \cite{Parrot} and Pie \cite{Pie} expose inter-request dependencies for optimizing multi-turn LLM applications. LAPS \cite{LAPS} observes that mixed prefill lengths within a batch create compute skew and disaggregates requests by prompt length. Continuum \cite{Continuum} directly targets agentic workloads by introducing a KV cache time-to-live policy that retains state across tool calls, avoiding redundant recomputation. All four systems assume homogeneous GPU pools and do not exploit hardware heterogeneity in their placement or routing decisions. HexGen-Flow \cite{HexGen-Flow} is the closest to our setting, dispatching multi-stage agentic workflows to heterogeneous GPUs. However, its cost model assumes a fixed Text-to-SQL pipeline and is not applicable to general agentic workloads.

To our knowledge, no existing system jointly exploits GPU heterogeneity and the multi-turn structure of agentic workloads---specifically, the asymmetry between a long first-turn prefill and short subsequent-turn incremental prefills---to adaptively route computation and minimize end-to-end job completion time. We address this gap in \name{}
}

\newedit{
Several recent systems serve LLMs on heterogeneous GPU clusters. Some formulate GPU allocation as constrained optimization over cost or throughput, assigning GPU types to model replicas or partitions \cite{Melange, Helix, HexGen}. Others route the prefill and decode phases onto distinct GPU tiers, matching the compute-bound prefill and memory-bound decode to hardware suited to each \cite{Splitwise, HexGen-2, ThunderServe, greenllm}. Although these approaches differ in optimization formulation and partitioning strategy, they share a common insight: LLM inference places non-uniform demands on compute, memory capacity, and memory bandwidth, making it profitable to map workload to GPU capabilities rather than deploying a single homogeneous GPU type. All of this work, however, targets single-turn inference. At conversation granularity, agentic workloads exhibit a stable two-phase structure, a compute-bound first-turn prefill followed by a memory-bound trajectory, which remains unexploited.
}

\section{Characterization of Agentic Workloads} \label{sec:characterization}
\begin{figure}[ht]
    \begin{center}
        \includegraphics[width=\columnwidth]{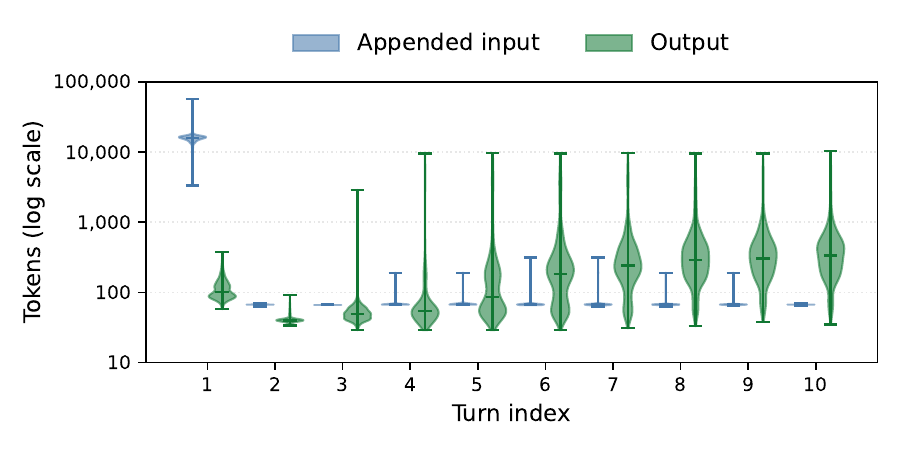}
    \end{center}
\vspace{-10pt}
    \caption{Input/Output token distribution of the first 10 turns from agentic traces. For the appended input, turn-1 is the input prompt, turn-2+ is the tool call response.}
    \label{fig:agentic-workload}
\end{figure}




While LLM and agentic workloads share similar computational phases, agentic workloads behave differently due to their underlying characteristics, as shown empirically in Figure \ref{fig:agentic-workload}. The figure depicts the input and output token distributions of agentic traces generated from SWE-bench\_bm25\_13K with swe-agent, using Qwen3-Coder-30B-A3B-Instruct as the trace-generation model. Agentic conversations begin with an initial input of tens of thousands of tokens, whereas prompts appended in later turns are far smaller, on the order of hundreds. Output token counts, in contrast, are unpredictable and exhibit large variance. To build the quantitative foundation to understand the impact of serving techniques on agentic conversations, we profile agentic workloads across computational phases, inputs tokens, context lengths, and GPU power budgets and analyze the impact of each variable on agentic serving.

\comment{
\myparagraph{Preliminaries}
All transformer-based LLMs, regardless of internal architectural choices such as mixture-of-experts layers or grouped-query attention, share the same two-phase inference structure: a prefill phase that computes KV cache entries for all input tokens, and a decode phase that generates output tokens autoregressively. Inference latency is bounded by the constraints and optimizations that impact each phase.
}

\myparagraph{Profiling Setup}
Experiments are conducted on NVIDIA A40 GPUs running Qwen3-0.6B in bf16 precision. This model size is chosen to leave sufficient GPU memory headroom to accommodate the large and growing KV caches of multi-turn agentic conversations, making the experimental setup representative of realistic agentic serving conditions. For heterogenous GPU experiments, we power cap GPUs at 2/3 of their Thermal Dynamic Power (TDP) to represent lower-power GPUs that has less compute capacity but similar memory hardware. While results are model- and hardware-specific, the general qualitative observations hold broadly across model sizes and GPU generations as the underlying computational properties are architectural rather than configuration-specific.

\subsection{Prefill Phase}
\label{sec:profiling-prefill}
We analyze the compute-bound prefill phase under multi-turn agentic conversations. We characterize each turn individually and use Time-to-first-token (TTFT) -- latency from request arrival to the first generated token -- for comparison. Our results show a clear picture: (1) first turn prefill latency is predictably expensive, but (2) caching and reusing the KV cache could substantially improve prefill latency at later turns, and (3) KV cache transfer between model replicas is relatively inexpensive in agentic scenarios. These insights motivate the design of \name{} in \S\ref{sec:design}.

\begin{figure}[ht]
    \begin{center}
        \includegraphics[width=\columnwidth]{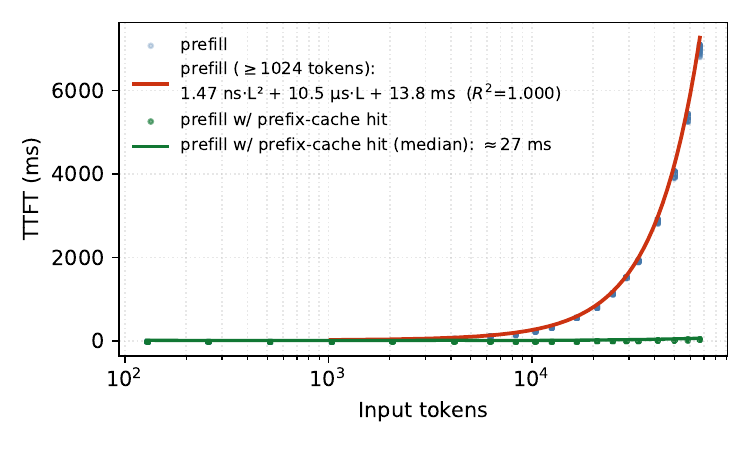}
    \end{center}
\vspace{-10pt}
    \caption{Characterization of TTFT under large inputs with and without caching. Uncached agentic prompts ($\ge10^4$) incur quadratic latency overheads but remain predictable as shown in the label ($R^2=1.0$). }
    \label{fig:prefill}
\end{figure}

\myparagraph{Initial prefills are expensive but predictable.}
We profile prefill latency across a range of input lengths; results are shown in Figure~\ref{fig:prefill}. TTFT is nearly constant for short prompts, but grows quadratically once attention becomes the dominant bottleneck. Prior work often models prefill latency as linear~\cite{databricks2023inference}, which is accurate when linear projection layers dominate the quadratic attention term. First turn agentic prompts, however, can reach tens of thousands of tokens, where attention is no longer negligible and a quadratic fit is substantially more accurate. In both regimes, TTFT is determined primarily by input length, allowing prefill latency to be estimated accurately from the input-token count alone. \name{} uses this observation to route incoming conversations away from saturated prefill workers before they incur queueing delay.

\myparagraph{Prefix caching reduces TTFT to near-constant in multi-turn agentic workloads} Figure \ref{fig:prefill} includes the latency of prefill stage with prefix-caching across a range of input tokens. As shown, prefix caching reduces TTFT to near-constant, reducing latency by two orders of magnitude under long inputs. Since relatively small prompts are appended to the whole history context after the first turn, prefix caching prior prompts and only calculating new KV cache entries for the appended tokens would substantially speed up the prefill stage at later turns.

\begin{figure}[ht]
    \begin{center}
        \includegraphics[width=\columnwidth]{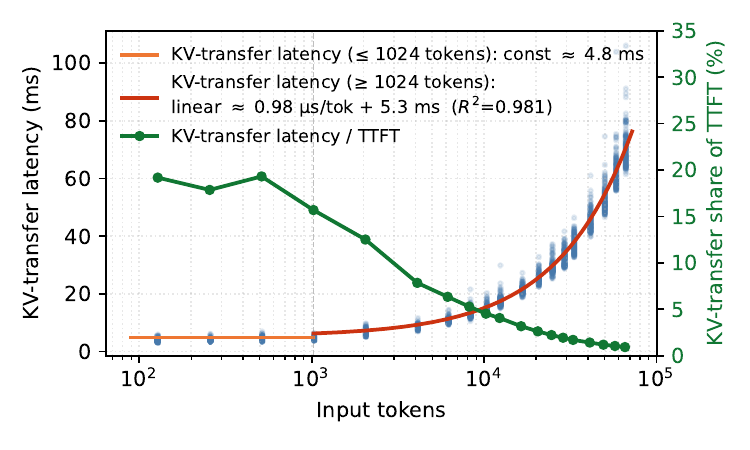}
    \end{center}
\vspace{-10pt}
    \caption{KV-transfer overhead is constant at short input lengths ($<1024$) and scales linearly at long inputs ($\ge1024$). 
    KV-transfer overhead between GPUs is significant when input is short but is dominated by Prefill latency in agentic workloads where inputs are tens of thousands of tokens long.}
    \label{fig:disagg}
\end{figure}

\myparagraph{KV cache transfer between GPUs is marginal in agentic workloads} Prefill-decode disaggregation -- \ie, when prefill and decode phases occur on separate model replicas -- incurs additional KV cache transfer between the prefill and decode GPU. Figure \ref{fig:disagg} depicts the latency overhead of KV-transfer and its relative impact on TTFT across a range of input tokens. While KV-transfer contributes significantly to TTFT at low input tokens ($>20\%$), it scales linearly with respect to input tokens and its impact is thus overshadowed by quadratic prefill latency at long inputs, precisely the operating domain of agentic workloads. Moreover, if no disaggregation happens for the rest of the turns given huge speedup by prefix caching, the KV-transfer overhead becomes negligible. \name{} relies on this insight when scheduling the first turn prefill stage for agentic requests.

\subsection{Decode Phase}
\label{sec:profiling-decode}
Output tokens are generated autoregressively across many iterations throughout the decode stage. We conclude that (1) agentic workloads are memory bound, (2) prefix-caching reduces collocation overheads, and (3) decoding latency is unpredictable. Our conclusions inform the the design of \name{} in \S\ref{sec:design}.


\begin{figure}[ht]
    \begin{center}
        \includegraphics[width=\columnwidth]{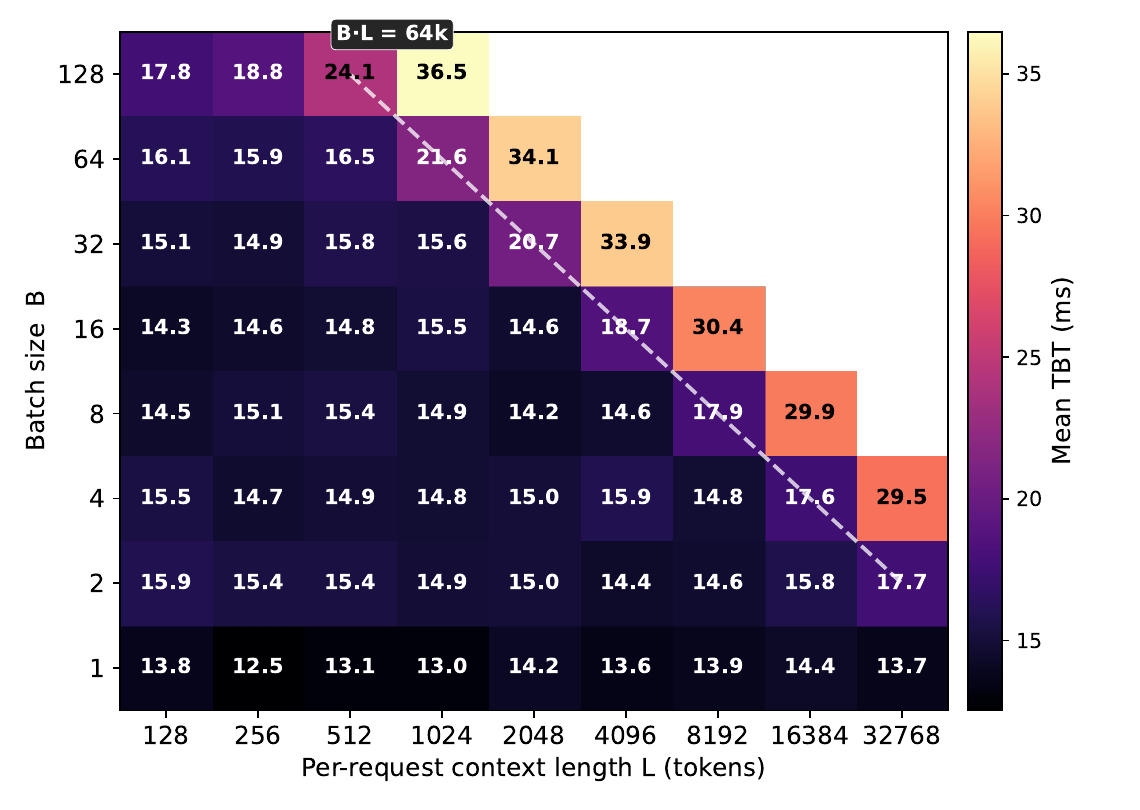}
    \end{center}
\vspace{-10pt}
    \caption{Heat map of mean time between tokens (TBT) across a range of batch sizes and context lengths. The dashed white line highlights the boundary between unsaturated (lower left) and saturated (upper right) memory bandwidth regions. Given the long context length of agentic tasks, agentic workloads lay in the saturated memory bandwidth domain.}
    \label{fig:decode_saturate}
\end{figure}

\myparagraph{Agentic workloads are memory-bound} We profile the average TBT across a range of batch sizes and KV lengths. As shown in Figure \ref{fig:decode_saturate}, mean TBT is relatively stable in low memory configurations but increases substantially under high batch sizes or long contexts lengths where memory bandwidth is saturated. Given the long context length of agentic tasks, these workloads are memory-bound during the decode stage.

\begin{figure}[ht]
    \begin{center}
        \includegraphics[width=\columnwidth]{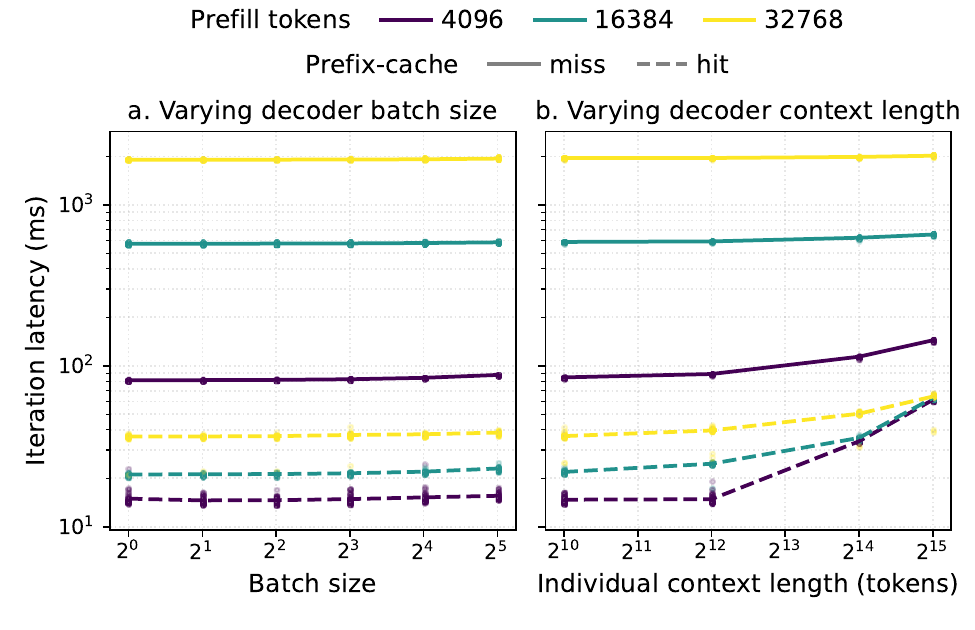}
    \end{center}
\vspace{-10pt}
    \caption{Iteration latency of collocated prefill and decode stages. (a) Latency of a prefill request arriving after 58 -- 92 decode iterations across a range decoder batch sizes. (b) Latency of a prefill request arriving after a range of decoding iterations with a fixed batch size (8). Prefix-caching significantly improves collocation overhead.}
    \label{fig:collocation}
\end{figure}

\comment{
\myparagraph{Prefix-caching significantly reduces collocation overheads} We conduct a series of experiments to understand the impact of collocating prefill and decode stages on a per iteration basis. In the first experiment, we begin a number of decode tasks and then introduce a prefill request after 58 -- 92 decode iterations and measure the collocated iterations latency. Figure \ref{fig:collocation}.a depicts these results across a range of initial decode tasks (batch size) and prefill size requests. In our second experiment, we maintain a consistent batch size and introduce the prefill request after a range of iterations. Figure \ref{fig:collocation}.b depicts these results across a range of prefill size requests. Both figures depict results with and without prefix-cache hits. As shown, prefix-caching significantly improves results by approximately one order of magnitude in both experiments; although, improvements are limited in large context lengths (\ie, the operating regime of agentic workloads) where memory bandwidth is saturated. Hence, \name{} utilizes prefix caching to reduce collocation overheads. \jerry{needs rework}
}

\newedit{
\myparagraph{Prefix-caching changes what impacts collocation overheads} We conduct a series of experiments to understand the impact of collocating prefill and decode stages on a per iteration basis. In the first experiment, we begin a number of decode tasks and then introduce a prefill request after 58 -- 92 decode iterations and measure the collocated iterations latency. Figure \ref{fig:collocation}.a depicts these results across a range of initial decode tasks (batch size) and prefill size requests. In our second experiment, we maintain a consistent batch size and introduce the prefill request after a range of iterations. Figure \ref{fig:collocation}.b depicts these results across a range of prefill size requests. Both figures depict results with and without prefix-cache hits. As shown, prefix-caching significantly improves results by approximately one order of magnitude in both experiments. However, when the decoding context gets larger to where memory cost exceeds to prefill compute cost, the iteration latency is largely affected by the decoding context length. At 262,144 active kv-cache tokens, the iteration latency is determined by the decoding instead of the collocated prefill. Hence, any prediction of such collocated iteration is invalid if the active kv-cache usage is not taken into account.
}

\begin{figure}[ht]
    \begin{center}
        \includegraphics[width=\columnwidth]{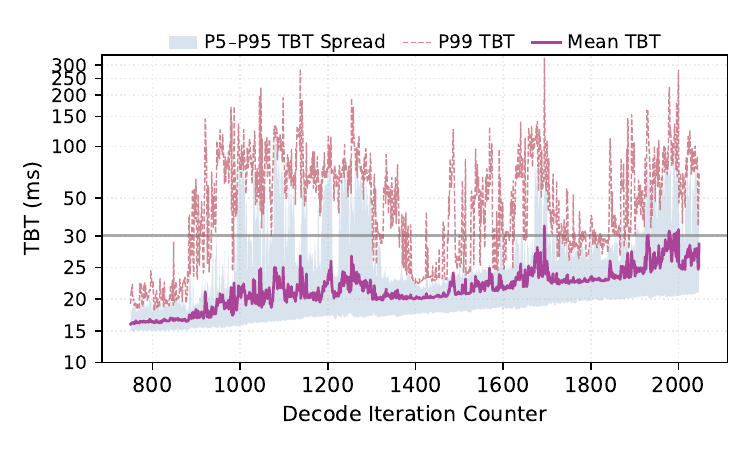}
    \end{center}
\vspace{-10pt}
    \caption{High variance in TBT throughout a long decode}
    \label{fig:decode_variable}
\end{figure}

\myparagraph{Decoding latency is highly variable} 
Our experimental results indicate that decoding latency is highly variable, and unlike the prefill stage, scheduling decisions should not rely on estimating decoding latency. 
We record the TBT for 64 different, batched prompts across 32 runs 
and report the iteration-level results in Figure \ref{fig:decode_variable}. While the mean TBT generally increases as the context length and number of iterations increase, individual TBT remains unpredictable. Moreover, as shown in Figure \ref{fig:collocation}.b, iteration latency under collocation and prefix-caching depends on the prefill input size and the context length of running decode tasks. Finally, the total number of output tokens is input-dependent and cannot be known apriori (Figure \ref{fig:agentic-workload}. Overall,  while decoding latency is correlated with memory bandwidth utilization, we find that accurately estimating the per iteration or end-to-end decoding latency is a challenging task that depends on many complex or unknown factors. Hence, \name{}'s design opts to not rely on such estimates for scheduling purposes.

\subsection{Heterogeneous Serving}
\label{sec:profiling-het}
We vary the GPU power caps to simulate the impact of heterogeneous agentic serving on the prefill and decode phase. We find that prefill and decode are impacted significantly and marginally, respectively. These findings further motivate \name{}'s design choices in \S\ref{sec:design}.

\begin{figure}[ht]
    \begin{center}
        \includegraphics[width=\columnwidth]{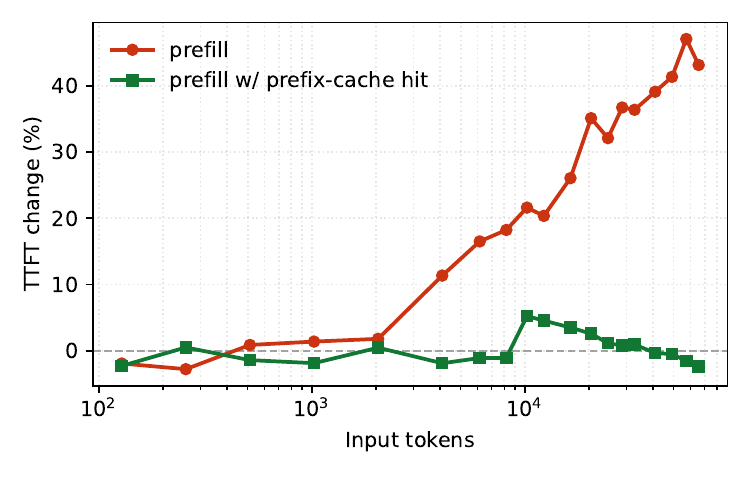}
    \end{center}
\vspace{-10pt}
    \caption{TTFT difference between uncapped GPUs and GPUS capped at 2/3 TDP. Power capping impacts the prefill stage heavily but has a marginal effect on prefix caching. }
    \label{fig:cap-prefill}
\end{figure}

\myparagraph{Power capping impacts the prefill stage heavily but has a marginal effect on prefix caching} We repeat the experiments in Figure \ref{fig:prefill} on power-capped GPUs and show the difference in performance in Figure \ref{fig:cap-prefill}. Since prefill is compute-bound at high input token ranges, power-capping significantly increases TTFT. On the other hand, prefill with prefix-caching remains relatively stable irrespective of input size. This insight harmonizes well with \name{}'s heterogeneous mapping: turn 1 prefill stages are scheduled on the best available GPUs, whereas later prefill stages with prefix-cache hits can occur on less powerful decode GPUs with marginal performance penalties.

\begin{figure}[ht]
    \begin{center}
        \includegraphics[width=\columnwidth]{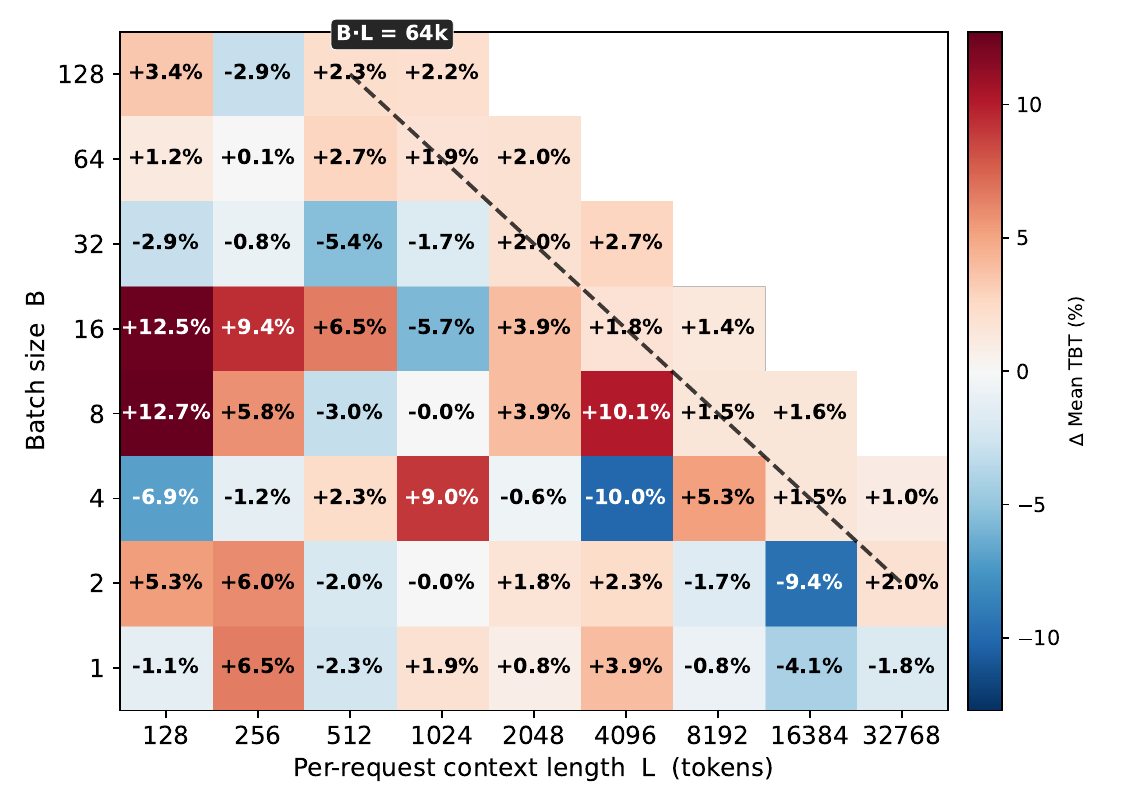}
    \end{center}
\vspace{-10pt}
    \caption{Heat map of mean TBT difference uncapped GPUs and GPUS capped at 2/3 TDP across a range of batch sizes and context lengths. The dashed black line highlights the boundary between unsaturated (lower left) and saturated (upper right) memory bandwidth regions. Power capping has a marginal effect in the saturated memory bandwidth domain.}
    \label{fig:cap-decode}
\end{figure}

\myparagraph{Power capping has a marginal effect on the decode stage of agentic workloads} We repeat the experiments in Figure \ref{fig:decode_saturate} on power-capped GPUs and show the results in Figure \ref{fig:cap-decode}. As depicted, decoding is marginally impacted when memory is saturated, \ie, the operating regime where agentic operations occur. This insight allows \name{} to delegate decoding to less powerful GPUs when possible.

\section{\name{}}
\label{sec:design}

\begin{figure}[ht]
    \begin{center}
        \includegraphics[width=\columnwidth]{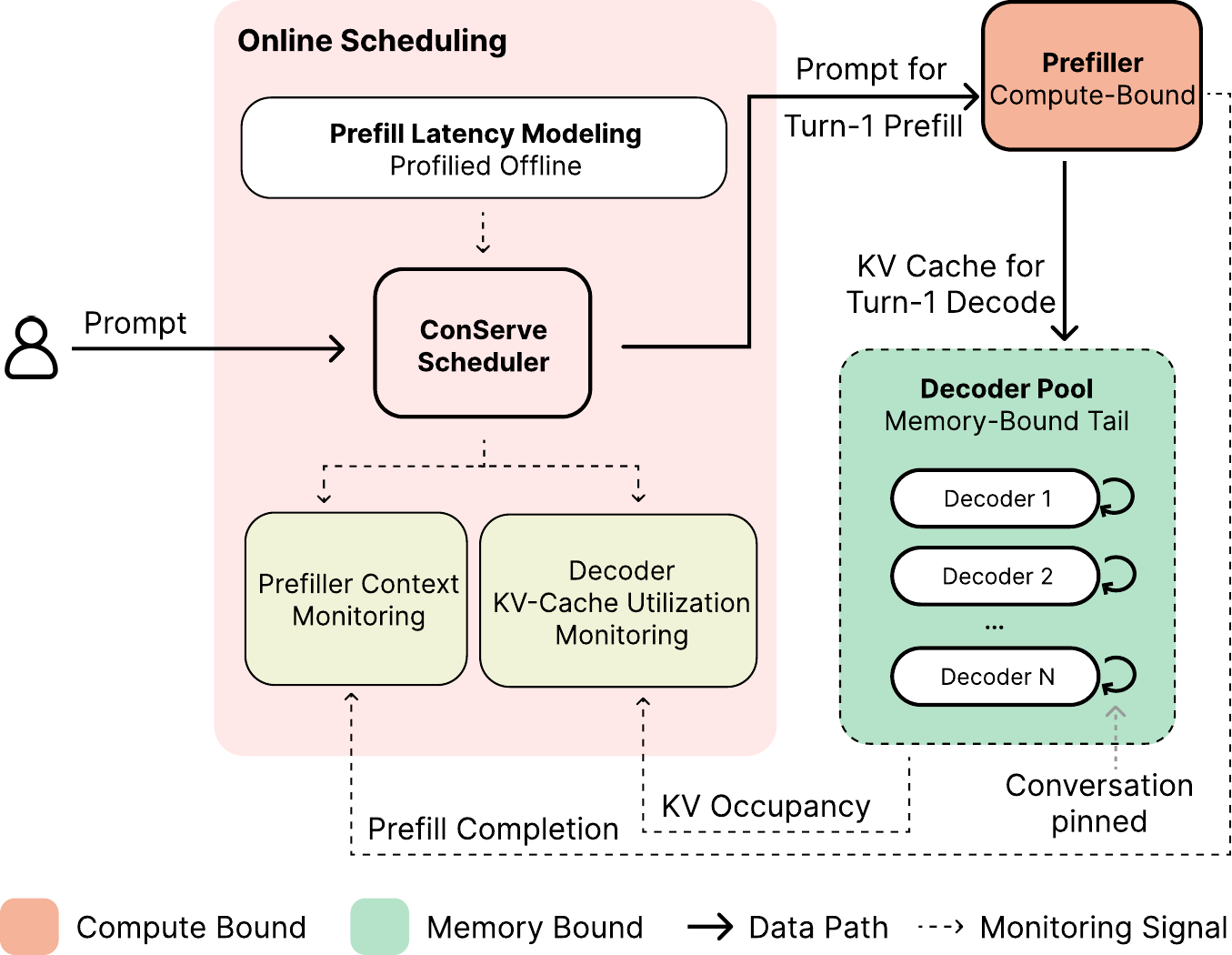}
    \end{center}
\vspace{-10pt}
    \caption{\name{} System Architecture.}
    \label{fig:design}
\end{figure}

\systemname{} instantiates conversation-level scheduling with a minimal placement policy. As illustrated in Figure \ref{fig:design}, the system runs on a prefill-decode disaggregated architecture: one model replica is dedicated to processing input prompts (the \emph{prefiller node}), and the remaining replicas are dedicated to generating output tokens (the \emph{decoder replicas}). A conversation is bound to one decoder replica at arrival, and the binding persists for the conversation's lifetime. The first-turn prefill is routed to the prefiller where the long initial prompt is processed; the resulting KV cache is transferred once to the bound decoder. The first-turn decode and every subsequent turn's incremental prefill and decoding then execute on that decoder with full KV cache reuse across turns. The system makes one routing decision per conversation and predicts no decode-side quantity at any point.

A \systemname{} deployment consists of one prefill node and $N$ decoder replicas, with $N$ chosen such that the prefill node saturates before the decoder pool under the workload's input-to-output token ratio. This deliberate over-provisioning of decoders places the system's throughput ceiling on the prefill side, where the relevant signal (input token rate) is observable at admission time and the cost-per-token relationship is deterministic (§\ref{sec:profiling-prefill}). The scheduler can therefore decide whether to admit a new conversation, where to bind it, and when to scale out by reading current system state rather than forecasting decode behavior.

\subsection{Instance Configuration}
\label{sec:design-config}

Provisioning $N$ is bounded by two constraints on the decoder side. The first is throughput: aggregate decoder throughput must process the tokens consumed and generated across each conversation's lifetime. The second is memory: each decoder pins its bound conversations and holds their KV cache in GPU memory, limiting concurrent conversations per replica to $B$ slots, where $B$ is determined by GPU memory capacity and per-conversation peak KV footprint.

Let $R$ denote the arrival rate, $T_d$ the per-decoder token throughput, $L_d$ the mean per-conversation token volume handled by decoders (turn-1 decode plus all turn 2+ prefill and decode), and $W$ the mean wall-clock lifetime including time spent on external tool calls. $N$ decoders must satisfy both constraints simultaneously:

\begin{align}
N \cdot T_d &\geq R \cdot L_d \quad \text{(throughput)} \\
N \cdot B &\geq R \cdot W \quad \text{(memory)}
\end{align}

The prefill node saturates at arrival rate $R^* = T_p / L_{in}$, where $T_p$ is the prefill input throughput and $L_{in}$ the mean first-turn input length. \systemname{} provisions $N$ as an integer more than satisfying both inequalities at $R = R^*$. This places the throughput bottleneck on the prefill node, where input token rate maps deterministically to expected utilization via the offline-profiled latency curve. The specific value of $N$ used in our evaluation is given in §\ref{sec:eval-setup}.

\subsection{Reactive Scheduling}
\label{sec:design-scheduling}
The scheduler operates on two signals, both directly observable. The first, the prefill latency curve, is profiled offline as a deterministic function of input token count (§\ref{sec:profiling-prefill}); given an incoming conversation's first-turn prompt length, the scheduler reads off the prefill node's expected utilization in constant time. The second is per-decoder \emph{active} KV cache occupancy. Because every turn reuses the conversation's accumulated history, every byte of KV cache held for a live conversation is part of its active working set; the scheduler decrements occupancy on conversation termination so the signal always reflects only currently active state rather than allocated-but-idle memory. Both signals are properties of state the system already maintains, not forecasts of what the workload will do next.

A new conversation enters the system at its first turn. The first-turn prefill is routed unconditionally to the prefill node. Once the prefill completes and the KV cache is transferred, the scheduler binds the conversation to the decoder replica with the lowest current KV cache occupancy. The binding is persistent: every subsequent turn for this conversation executes on the same decoder, with no re-evaluation of where it should go.

Because the two signals are direct measurements of system state, the same observability extends to capacity management. Imminent prefill saturation is detectable from the input token rate against the profiled curve, and decoder saturation surfaces as aggregate KV cache pressure across the pool. Provisioning additional capacity in either tier follows from the same observations the scheduler already makes.

\subsection{Heterogeneous GPU Mapping}
\label{sec:design-het}

\systemname{} segments a conversation into a compute-bound first-turn prefill and a memory-bound tail, mirroring the phase structure that prefill-decode disaggregation exploits. This segmentation maps cleanly onto heterogeneous GPU tiers. Memory-bound workloads scale less aggressively with newer hardware than compute-bound workloads do, so decoder replicas, which run the memory-bound tail, can execute on older lower-power GPUs without proportional throughput loss, while the prefill node takes the newest high-power GPU available.

This mapping requires no changes to the scheduling logic. The same two observable signals drive placement on either uniform or heterogeneous hardware, and the conversation-level commitment ensures KV state is transferred between tiers at most once per conversation. Energy efficiency emerges as a consequence: the system spends high-power compute only on the turn that requires it, while the memory-bound tail runs on hardware that consumes substantially less power.

\section{Evaluation}
Our evaluation addresses the following questions:
\begin{itemize}[leftmargin=2em]
    \item \textbf{Q1:} Does \name{} match or beat all baselines on latency?
    \item \textbf{Q2:} Does \name{} maintain SLO at saturation?
    \item \textbf{Q3:} Does \name{} avoid the routing error sensitivities of per-turn approaches?
    \item \textbf{Q4:} Does the heterogeneous variant translate to energy efficiency wins?
\end{itemize}

\begin{figure} [t]
    \begin{center}
        \includegraphics[width=\columnwidth]{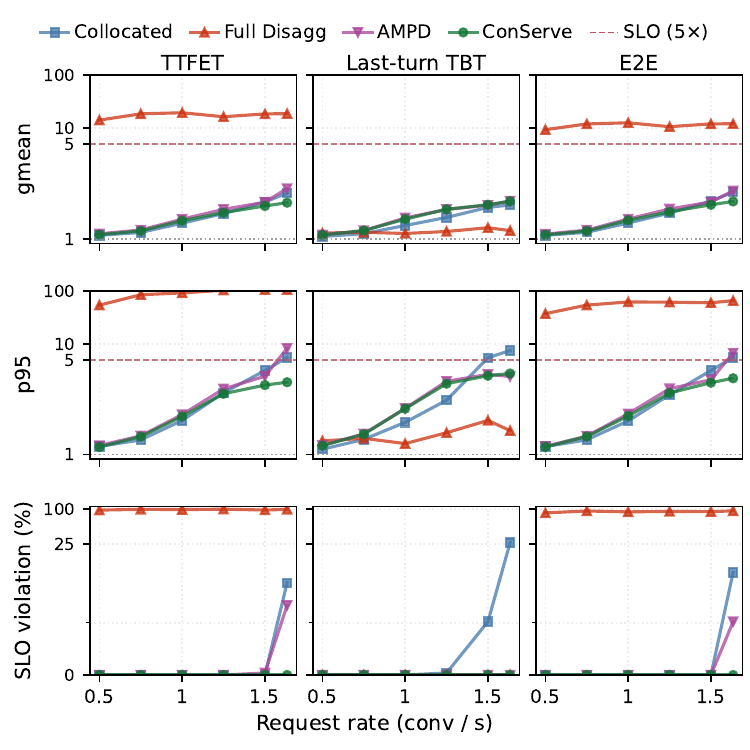}
    \end{center}
    \vspace{-5pt}
    \caption{Normalized agentic performance results over multiple request arrival rates (lower is better). AMPD has 10\% wrong prediction rate.}
    \label{fig:headline}
    \vspace{-3.5pt}
\end{figure}

\begin{figure} [t]
    \begin{center}
        \includegraphics[width=\columnwidth]{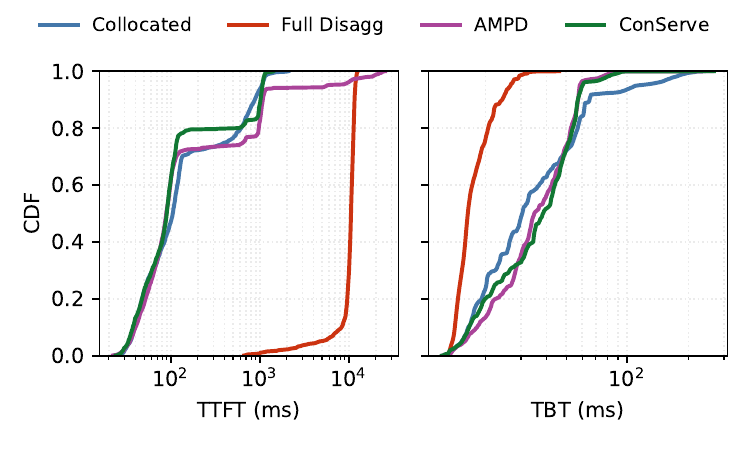}
    \end{center}
    \vspace{-5pt}
    \caption{Overall TTFT and TBT distribution at RPS 1.634.}
    \label{fig:ttft_tbt}
    \vspace{-3.5pt}
\end{figure}

\subsection{Experiment Setup}
\label{sec:eval-setup}

\paragraph{Workload and hardware} We evaluate on agentic traces generated from SWE-bench\_bm25\_13K with swe-agent, using Qwen3-Coder-30B-A3B-Instruct as the trace-generation model. Traces are replayed on a 4-GPU NVIDIA A40 machine with Qwen3-0.6B as the served model. The smaller served model leaves sufficient memory headroom for the long KV caches characteristic of multi-turn agentic conversations. To simulate the serving performance on heterogeneous GPUs, we power cap 3 GPUs to 2/3 of their TDP (300W) to represent prior-generation GPUs with less compute capacity but similar memory hardware.

\myparagraph{System configuration} Each served model replica occupies one GPU. The prefiller sustains roughly 25k input tokens per second, while each decoder produces 1k output tokens per second and accommodates approximately 300k tokens of KV cache. We allocate one prefiller and three decoders across the four-GPU machine. As a sanity check following the analysis in §\ref{sec:design-config}, an average of 15k input tokens and 1k output tokens per conversation would require at least 1.67 decoders per prefiller. The 3 decoder configuration is a guarantee that the prefiller saturates before the decoders.

\myparagraph{Implementation and Baselines}

We implement \name{} on with vLLM \cite{vLLM} as the LLM serving engine and LMCache \cite{cheng2025lmcache} as the Prefill-Decode disaggregation manager. We configure the prefiller model replica with a vLLM standard 8192 chunked token size per batch to prevent aggressive queueing due to long context. We compare against three baselines. 
\begin{itemize}[leftmargin=2em]
    \item \textbf{Collocated} uses all four GPUs as mixed-batch replicas, and runs each conversation on a single replica with prefill and decode batched together. Each model replica is configured with a 2944 chunked token size. This size is derived from our offline profiling, such that the iteration latency is within 5x baseline SLO, as practiced in prior work \cite{dynamoLLM}.
    \item \textbf{Full Disaggregation} routes every turn through a dedicated prefill node. 
    \item We also implement \textbf{AMPD} \cite{ampd} at our best effort, which performs per-turn prediction-based disaggregation for multi-turn workloads. As bi-directional KV cache transfer (between two model replicas) and management is infeasible to implement in a reasonable amount of time, we simulate the latency of KV cache transfer and stall requests for that period. We further post-process the experiment trace with our offline profiling results at best effort. Because in all our experiments, executing the turn 2+ prefill locally on the decoders is more effecient than migrating it to the prefiller, all reported AMPD results except Figure \ref{fig:wrong_prediction} induce a 10\% wrong prediction rate, i.e., the scheduler migrates the turn 2+ prefill to the prefiller for 10\% of all turns.
\end{itemize}

\myparagraph{Metrics} We report three latency metrics: TTFET, last-turn TBT, and end-to-end (E2E) latency, motivated in §\ref{sec:bg-agentic}. Together they capture the three user-facing performance dimensions that matter for multi-turn agentic conversations. For comparison with prior work, we additionally report conventional per-turn TTFT and TBT distributions, though these metrics conflate intermediate tool-call turns with the final user-facing reply and therefore do not reflect end-user experience.

\subsection{Q1: Agentic Conversation Performance}
\myparagraph{Agentic performance} Figure~\ref{fig:headline} reports normalized geometric mean and p95 latencies for TTFET, last-turn TBT, and E2E across the five systems at request rates from 0.5 to 1.5 conversations per second under Poisson arrivals. We additionally report results at 1.634 conv/s, which is not a Poisson rate but a synthesized arrival pattern that holds the prefiller exactly at its saturation throughput, isolating behavior at the system's capacity limit. For AMPD we fixed a wrong prediction rate of 10\%, with further discussion in section \ref{sec:q4}.

\name{} outperforms AMPD by reducing up to 19.17\% and +51.08\% on geometric mean and p95 TTFET. At low to moderate loads (0.5 to 1.25 conv/s), \name{}, AMPD, and Collocated remain within 2$\times$ of the baseline. At saturation (1.5 to 1.634 conv/s), \name{} holds steady while Collocated and AMPD degrade sharply at p95, reflecting their respective structural failure modes: Collocated suffers from prefill-decode contention growing with load, and AMPD accumulates routing-decision errors that compound into prefiller queue pressure (§\ref{sec:q4}). Full Disaggregation is uncompetitive at every operating point, with gmean TTFET and E2E exceeding 10$\times$ the baseline. Routing every turn through the prefill node forfeits the cross-turn KV cache reuse that dominates efficiency in multi-turn workloads, and additionally pays a per-turn KV transfer cost.

Last-turn TBT inverts the ordering. Full Disaggregation achieves the lowest TBT across all loads, since its decoder replicas process only decode tokens with no prefill interference. At saturation, \name{}'s gmean last-turn TBT (2.49$\times$ baseline) is 91.85\% higher than Full Disaggregation's (1.35$\times$). This penalty is the price of preserving cross-turn KV cache reuse, and is dwarfed by the prefill latency that Full Disaggregation incurs on every turn: its gmean TTFET reaches 18.78$\times$ baseline against \name{}'s 2.53$\times$, translating to 4.7$\times$ worse E2E. AMPD produces essentially identical TBT to \name{} despite disaggregating roughly 10\% of turn 2+ prefills, showing that partial disaggregation at this rate gains no measurable TBT benefit while still paying the TTFET cost of those routing decisions (§\ref{sec:q4}). Collocated has the worst p95 TBT at saturation (7.52$\times$ baseline, 72\% worse than \name{}), reflecting the prefill-decode contention from batching long initial prefills with ongoing decoder workloads.

\myparagraph{A deeper look at LLM metrics} Figure~\ref{fig:ttft_tbt} shows the per-turn TTFT and TBT CDFs across all turns and conversations. Full Disaggregation pays both the full prefill cost and the KV transfer on every turn, producing the rightmost TTFT distribution; its prefill-free decoders produce the leftmost TBT distribution.

The TTFT CDF resolves into three regimes for the remaining systems. Below the 0.7 percentile, \name{}, AMPD, and Collocated overlap: these are turn 2+ prefills, which complete quickly regardless of where they execute. Between 0.7 and 0.8, \name{} pulls ahead significantly. This regime contains turn-1 prefills, which in Collocated batch with ongoing turn 2+ prefills on the same replicas, and in AMPD wait behind the queue formed by wrongly disaggregated turn 2+ prefills on the prefill node. Above 0.8, the ordering inverts: Collocated becomes best because its four-GPU pool prevents compute saturation, while AMPD becomes worst as its accumulated prefill queue stretches into the tail.

The TBT CDF completes the picture. \name{} and AMPD produce nearly identical distributions across the full range; AMPD's 10\% per-turn disaggregation does not meaningfully change which prefills reach the decoders. Below the 0.7 percentile, Collocated outperforms both because it dedicates four GPUs to decoding while \name{} and AMPD dedicate only three, giving Collocated one-third more aggregate memory bandwidth. Above the 0.7 percentile, this advantage inverts as long turn-1 prefills batch with ongoing decode on Collocated's replicas and inflate per-step latency, while \name{} and AMPD insulate the decoders from this contention.

\subsection{Q2: Serving Throughput and Quality}
\label{sec:q2}

Figure~\ref{fig:headline} (third row) reports SLO violation rates. We set the SLO threshold at 5$\times$ the baseline latency for each metric, where the baseline is measured by executing a single request without batching or interference. This 5$\times$ threshold follows standard practice in prior LLM serving work. \name{} sustains zero SLO violations across all three metrics at every request rate tested, including the 1.634 conv/s saturation point. The baselines fail at saturation in metric-specific patterns that reveal their underlying failure modes. Collocated violates both TTFET and last-turn TBT SLOs at 1.634 conv/s, since prefill-decode contention degrades both phases simultaneously. AMPD violates only the TTFET SLO, because its per-turn routing errors accumulate on the prefiller but leave the decoder workload unchanged (§\ref{sec:q4}). Full Disaggregation violates the TTFET and E2E SLOs at nearly 100\% across all loads, with last-turn TBT remaining under the SLO due to its interference-free decoder design. 

The structural reason for \name{}'s SLO robustness is that the prefiller's load is bounded by design. Only turn-1 prefills route to the prefill node, and the rate at which they arrive is gated by incoming conversation arrival. Since turn-1 prefill latency is a deterministic function of input token count (§\ref{sec:profiling-prefill}), the system operates exactly at the prefiller's saturation capacity without exceeding it. Neither baseline has this property: Collocated absorbs all load on a single batched replica pool, and AMPD's per-turn routing injects unpredictable additional prefiller load that breaks the operating bound.

\subsection{Q3: Pitfalls of Per-Turn Routing}
\label{sec:q4}
\begin{figure} [t]
    \begin{center}
        \includegraphics[width=\columnwidth]{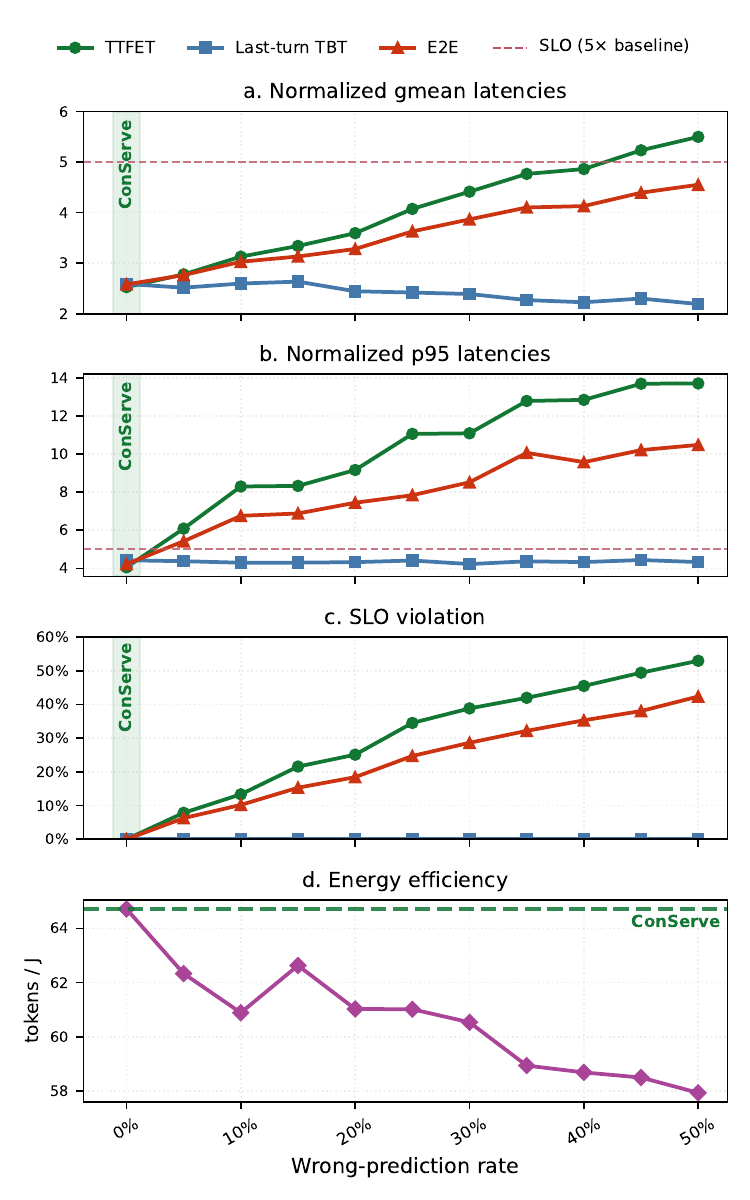}
    \end{center}
    \vspace{-5pt}
    \caption{Comparison between \name{} and AMPD over different wrong prediction rate.}
    \label{fig:wrong_prediction}
    \vspace{-3.5pt}
\end{figure}

AMPD's per-turn predictions are susceptible to three sources of error: the cost model does not account for decoder KV cache utilization, which directly affects decode throughput (§\ref{sec:profiling-decode}); when incremental prefills batch with heavy ongoing decode, the resulting latency has high variance that offline profiling cannot capture; and most consequentially, the cost model does not account for prefiller queueing pressure, so each erroneously disaggregated turn 2+ prefill adds to the prefiller's queue and delays all subsequent prefill workload. \name{} sidesteps all three by scheduling on incoming input prompt size alone, which is fully observable and makes the prefiller's saturation point deterministic, enabling predictable autoscaling.

Figure~\ref{fig:wrong_prediction} shows that AMPD is highly sensitive to even small cost-estimation errors. \newedit{At a 0\% wrong-prediction rate, AMPD routes every turn 2+ prefill locally and reduces to \name{}, so the curves begin at \name{}'s operating point.} At just 5\% wrong predictions, SLO violations reach roughly 7.8\% for TTFET and 6.3\% for E2E. As the wrong-prediction rate climbs to 50\%, both gmean and p95 latencies grow linearly and SLO violations exceed 50\%. The linearity follows directly from the queueing mechanism: each wrong decision adds one unit of unanticipated load to a system operating at saturation, so the resulting queue delay scales with the cumulative excess load. Our evaluation runs a finite trace subset, after which the prefiller queue drains. In a sustained production workload, the queueing pressure continues to accumulate, and the SLO violations observed here form a lower bound on AMPD's deployment-scale impact. Last-turn TBT remains flat throughout, since routing errors affect intermediate turns but never reach the user-visible final turn. \name{} has no equivalent degradation curve: it makes no per-turn routing decision and therefore generates no prediction error.

\newedit{The same routing errors degrade energy efficiency (Figure~\ref{fig:wrong_prediction}.d). A wrongly migrated turn 2+ prefill carries substantial reused KV cache; mixing this memory-heavy work with the compute-bound turn-1 prefills on the prefiller lowers the prefiller's utilization. The misrouted conversations also incur longer latency, and these delays accumulate into a longer wall-clock time to serve the full workload. At the 1.634 conv/s saturation point, with 10\% wrong predictions, AMPD takes 15\% longer than \name{} to drain the workload while delivering 7.5\% lower tokens-per-joule. Because the workload processes a fixed token volume, the lower efficiency reflects higher total energy; that the 15\% time penalty exceeds this energy increase indicates the GPUs spend the extra time under-utilized, idling on queue rather than computing. The effect compounds with error rate: AMPD's tokens-per-joule declines monotonically toward 58 tokens/J at a 50\% wrong-prediction rate, while \name{} holds constant by construction.}

\subsection{Q4: Serving with Heterogeneous GPUs}
\label{sec:q3}
\begin{figure} [t]
    \begin{center}
        \includegraphics[width=\columnwidth]{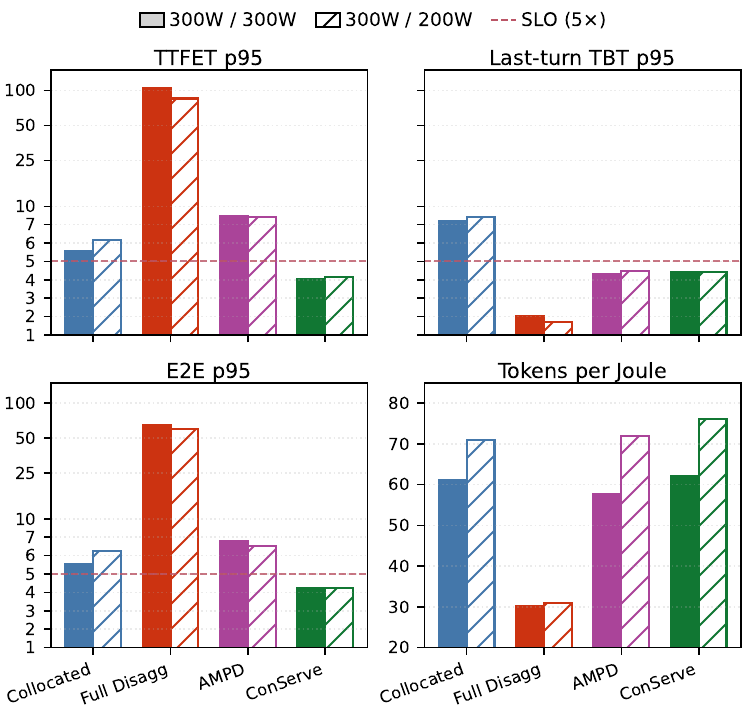}
    \end{center}
    \vspace{-5pt}
    \caption{P95 agentic performance and energy efficiency on heterogeneous GPUs.}
    \label{fig:eval-het}
    \vspace{-3.5pt}
\end{figure}

\comment{

{\color{red}
Figure \ref{fig:eval-het} compares all four systems under a homogeneous configuration ($300\,\text{W}/300\,\text{W}$) and a heterogeneous one ($300\,\text{W}/200\,\text{W}$), where the decoder-side GPUs are power-capped to simulate heterogeneity (§\ref{sec:eval-setup}).

All four systems are largely insensitive to the power cap on TTFET, last-turn TBT, and E2E geometric mean latency. Specifically, ConServe shows only an x\%, x\%, and x\% change in TTFET, last-turn TBT, and E2E geometric mean latency, respectively, remaining within the $5\times$ SLO on every metric. The other systems exhibit the same structural patterns observed earlier: Full Disaggregation violates the TTFET and E2E SLOs in both configurations, while its last-turn TBT stays lowest.

Figure \ref{fig:eval-het} (bottom right) shows tokens per Joule. ConServe achieves an x\% improvement under the heterogeneous configuration over its homogeneous one, the highest among all systems.
}

The structural reason is that decode is memory-bound (§\ref{sec:profiling-decode}): its throughput is gated by memory bandwidth, not compute. Power capping a GPU reduces compute capacity proportionally more than memory bandwidth, so the compute reduction is irrelevant to decode-dominated workloads. \name{}'s design routes all decode and turn 2+ prefills to the decoder replicas, which means the power cap targets exactly the operations that can absorb it. The heterogeneous proposition from §\ref{sec:design-het}, pairing newer high-power GPUs with older lower-power GPUs for prefill and decode respectively, holds empirically: the decode-side hardware can be downscaled in power without a measurable cost to user-visible latency.
}

\newedit{
\name{} segments each conversation into a compute-bound first-turn prefill and a memory-bound tail, mirroring the prefill-decode abstraction at conversation granularity. If this segmentation is faithful, the memory-bound tail should tolerate a reduced power budget on the decoder GPUs: its throughput is gated by memory bandwidth, not by the compute capacity a power cap removes. Figure~\ref{fig:eval-het} tests this, comparing a homogeneous configuration (300W prefill, 300W decoders) against a heterogeneous one (300W prefill, 200W decoders).

\name{} realizes two energy wins. Its scheduling alone is energy-efficient in the homogeneous configuration, since confining compute-bound work to one prefill node and running the memory-bound tail without per-turn disaggregation avoids the wasted transfers and idle cycles that lower the baselines' tokens-per-joule. Capping the decoders to 200W then adds a further gain: tokens-per-joule rises by 22.75\% while p95 TTFET, last-turn TBT, and E2E remain essentially unchanged. The memory-bound tail absorbs the power cap, so the heterogeneous mapping does not trade away \name{}'s scheduling advantage but compounds it.

The baselines do not share this property. Collocated batches prefill and decode on the same GPUs, so reducing decoder power to 200W starves the compute-bound prefill: its p95 TTFET worsens by 11.15\% under the heterogeneous configuration. Full Disaggregation cannot meet the SLO in either configuration, leaving its energy numbers moot. AMPD gains the same power-cap benefit as \name{}, since it also runs decode on the capped GPUs, but it carries the per-turn routing penalties characterized in §\ref{sec:q3}.
}

\section{Related Work}
\newedit{
\myparagraph{LLM Request Scheduling} 
LLM request scheduling has been studied along several axes. Iteration-level scheduling batches and preempts at the granularity of individual decode steps to raise utilization \cite{orca, llumnix}, and recent schedulers route or reorder requests using predicted output lengths or learned rankings \cite{fu-rank, jain-router}. A parallel line schedules across the prefill-decode split: disaggregation systems decide where each phase executes and how the two scale independently \cite{DistServe, Splitwise}, while intra-GPU and unified designs revisit where the split is drawn \cite{nexus, duetserve, taichi}. Elastic systems extend this to autoscaling prefill and decode capacity in response to load \cite{blitzscale, tokenscale}. Production frameworks fold all of these together: NVIDIA Dynamo unifies KV-cache-aware routing, conditional prefill-decode disaggregation, SLO-driven autoscaling, and tiered KV cache management in one serving system \cite{dynamo}. Whether these mechanisms are studied in isolation or integrated this way, the unit of scheduling remains the request, or its constituent phases and iterations. \systemname{} raises the unit to the conversation: it commits one placement per conversation and never revisits it, which is what lets scheduling condition on observable state rather than predicted per-request cost.

\myparagraph{KV Cache Management} 
A separate line of work targets how key-value state is stored and reused rather than where the computation that produces it runs. Hierarchical and distributed cache layers move KV state across memory tiers and across the cluster to extend reuse beyond a single instance \cite{cachedattention, mooncake, memserve, cheng2025lmcache}, and compression and offloading shrink or relocate the cache to fit larger working sets \cite{snapkv, cachegen}. These mechanisms operate at a different layer than \systemname{}, which schedules computation and assumes only standard local prefix caching beneath it. The two compose: richer cache management widens what can be reused, while \systemname{}'s conversation-level placement keeps the reusable prefix on the instance that will read it, so after the first turn it never crosses the network again.

\myparagraph{Agentic and Application-Aware Serving}
A growing body of work optimizes serving for the structure of agentic applications. Parrot recovers prompt sharing and request dependencies through application-level dataflow \cite{parrot}. A second line targets the tool-call period that separates an agent's turns: InferCept retains, discards, or swaps the paused KV cache while an external call runs \cite{infercept}, Continuum retains KV state across pauses under a time-to-live policy \cite{continuum}, and Sutradhara co-designs orchestrator and engine to overlap tool execution with subsequent prefill and decode \cite{sutradhara}. KVFlow evicts and prefetches cache entries according to an agent execution graph \cite{kvflow}, and Autellix raises the scheduling unit to the agentic program to order calls against head-of-line blocking \cite{autellix}. Several of these optimize conversation-level rather than per-turn objectives: Continuum targets job completion time, and Sutradhara optimizes final-answer latency, the same quantity \systemname{} captures as TTFET. All of them operate within a conversation whose physical placement is fixed, managing its cache, overlapping its tool calls, and ordering its requests on an instance already chosen. None makes the placement decision itself: where the first-turn prefill runs, when its KV state transfers, and which decoder holds the conversation for its lifetime. \systemname{} commits this decision once from observable state and is orthogonal to this line of work, which caches, overlaps, and schedules the calls within a conversation \systemname{} has already placed.

}

\section{Conclusion}
We introduce ConServe, a conversation-level disaggregation scheduler for agentic serving. ConServe raises the scheduling unit from the turn to the conversation, which collapses turn-level irregularity into a stable two-phase structure: a compute-bound first-turn prefill followed by a memory-bound tail. ConServe routes the first-turn prefill to a high-power prefiller provisioned as the system bottleneck, transfers the KV cache exactly once, and pins each conversation to a decoder replica for its lifetime, conditioning placement only on observable turn-1 input length and per-decoder KV occupancy rather than predicted per-turn cost. Experiments on agentic workloads show that ConServe reduces p95 TTFET by 51.08\% over per-turn routing while sustaining zero SLO violations at saturation, and its phase split maps onto heterogeneous GPU tiers for a further 22.75\% gain in energy efficiency.

\bibliographystyle{ACM-Reference-Format}
\bibliography{reference}


\begin{thebibliography}{50}


\ifx \showCODEN    \undefined \def \showCODEN     #1{\unskip}     \fi
\ifx \showISBNx    \undefined \def \showISBNx     #1{\unskip}     \fi
\ifx \showISBNxiii \undefined \def \showISBNxiii  #1{\unskip}     \fi
\ifx \showISSN     \undefined \def \showISSN      #1{\unskip}     \fi
\ifx \showLCCN     \undefined \def \showLCCN      #1{\unskip}     \fi
\ifx \shownote     \undefined \def \shownote      #1{#1}          \fi
\ifx \showarticletitle \undefined \def \showarticletitle #1{#1}   \fi
\ifx \showURL      \undefined \def \showURL       {\relax}        \fi
\providecommand\bibfield[2]{#2}
\providecommand\bibinfo[2]{#2}
\providecommand\natexlab[1]{#1}
\providecommand\showeprint[2][]{arXiv:#2}

\bibitem[Abhyankar et~al\mbox{.}(2024)]%
        {infercept}
\bibfield{author}{\bibinfo{person}{Reyna Abhyankar}, \bibinfo{person}{Zijian He}, \bibinfo{person}{Vikranth Srivatsa}, \bibinfo{person}{Hao Zhang}, {and} \bibinfo{person}{Yiying Zhang}.} \bibinfo{year}{2024}\natexlab{}.
\newblock \bibinfo{title}{InferCept: Efficient Intercept Support for Augmented Large Language Model Inference}.
\newblock
\showeprint[arxiv]{2402.01869}~[cs.LG]
\urldef\tempurl%
\url{https://arxiv.org/abs/2402.01869}
\showURL{%
\tempurl}


\bibitem[Agrawal et~al\mbox{.}(2024)]%
        {Sarathi-Serve}
\bibfield{author}{\bibinfo{person}{Amey Agrawal}, \bibinfo{person}{Nitin Kedia}, \bibinfo{person}{Ashish Panwar}, \bibinfo{person}{Jayashree Mohan}, \bibinfo{person}{Nipun Kwatra}, \bibinfo{person}{Bhargav~S. Gulavani}, \bibinfo{person}{Alexey Tumanov}, {and} \bibinfo{person}{Ramachandran Ramjee}.} \bibinfo{year}{2024}\natexlab{}.
\newblock \showarticletitle{Taming Throughput-Latency Tradeoff in {LLM} Inference with Sarathi-Serve}. In \bibinfo{booktitle}{\emph{18th {USENIX} Symposium on Operating Systems Design and Implementation, {OSDI} 2024, Santa Clara, CA, USA, July 10-12, 2024}}, \bibfield{editor}{\bibinfo{person}{Ada Gavrilovska} {and} \bibinfo{person}{Douglas~B. Terry}} (Eds.). \bibinfo{publisher}{{USENIX} Association}, \bibinfo{pages}{117--134}.
\newblock


\bibitem[{Anthropic}(2026)]%
        {anthropic_claude_code_limits_2026}
\bibfield{author}{\bibinfo{person}{{Anthropic}}.} \bibinfo{year}{2026}\natexlab{}.
\newblock \bibinfo{title}{Models, Usage, and Limits in Claude Code}.
\newblock \bibinfo{howpublished}{\url{https://support.claude.com/en/articles/14552983-models-usage-limits-in-claude-code}}.
\newblock
\newblock
\shownote{Accessed: 2026-05-30}.


\bibitem[Biswas et~al\mbox{.}(2026)]%
        {sutradhara}
\bibfield{author}{\bibinfo{person}{Anish Biswas}, \bibinfo{person}{Kanishk Goel}, \bibinfo{person}{Srivarshinee S}, \bibinfo{person}{Jayashree Mohan}, \bibinfo{person}{Alind Khare}, \bibinfo{person}{Anjaly Parayil}, \bibinfo{person}{Ramachandran Ramjee}, {and} \bibinfo{person}{Chetan Bansal}.} \bibinfo{year}{2026}\natexlab{}.
\newblock \bibinfo{title}{Sutradhara: An Intelligent Orchestrator-Engine Co-design for Tool-based Agentic Inference}.
\newblock
\showeprint[arxiv]{2601.12967}~[cs.DC]
\urldef\tempurl%
\url{https://arxiv.org/abs/2601.12967}
\showURL{%
\tempurl}


\bibitem[Chase(2022)]%
        {LangChain}
\bibfield{author}{\bibinfo{person}{Harrison Chase}.} \bibinfo{year}{2022}\natexlab{}.
\newblock \bibinfo{booktitle}{\emph{LangChain}}.
\newblock
\urldef\tempurl%
\url{https://github.com/langchain-ai/langchain}
\showURL{%
\tempurl}


\bibitem[Cheng et~al\mbox{.}(2025)]%
        {cheng2025lmcache}
\bibfield{author}{\bibinfo{person}{Yihua Cheng}, \bibinfo{person}{Yuhan Liu}, \bibinfo{person}{Jiayi Yao}, \bibinfo{person}{Yuwei An}, \bibinfo{person}{Xiaokun Chen}, \bibinfo{person}{Shaoting Feng}, \bibinfo{person}{Yuyang Huang}, \bibinfo{person}{Samuel Shen}, \bibinfo{person}{Kuntai Du}, {and} \bibinfo{person}{Junchen Jiang}.} \bibinfo{year}{2025}\natexlab{}.
\newblock \showarticletitle{LMCache: An Efficient KV Cache Layer for Enterprise-Scale LLM Inference}.
\newblock \bibinfo{journal}{\emph{arXiv preprint arXiv:2510.09665}} (\bibinfo{year}{2025}).
\newblock


\bibitem[Databricks(2023)]%
        {databricks2023inference}
\bibfield{author}{\bibinfo{person}{Databricks}.} \bibinfo{year}{2023}\natexlab{}.
\newblock \bibinfo{title}{LLM Inference Performance Engineering: Best Practices}.
\newblock \bibinfo{howpublished}{\url{https://www.databricks.com/blog/llm-inference-performance-engineering-best-practices}}.
\newblock


\bibitem[Fu et~al\mbox{.}(2024)]%
        {fu-rank}
\bibfield{author}{\bibinfo{person}{Yichao Fu}, \bibinfo{person}{Siqi Zhu}, \bibinfo{person}{Runlong Su}, \bibinfo{person}{Aurick Qiao}, \bibinfo{person}{Ion Stoica}, {and} \bibinfo{person}{Hao Zhang}.} \bibinfo{year}{2024}\natexlab{}.
\newblock \bibinfo{title}{Efficient LLM Scheduling by Learning to Rank}.
\newblock
\showeprint[arxiv]{2408.15792}~[cs.LG]
\urldef\tempurl%
\url{https://arxiv.org/abs/2408.15792}
\showURL{%
\tempurl}


\bibitem[Gao et~al\mbox{.}(2024)]%
        {cachedattention}
\bibfield{author}{\bibinfo{person}{Bin Gao}, \bibinfo{person}{Zhuomin He}, \bibinfo{person}{Puru Sharma}, \bibinfo{person}{Qingxuan Kang}, \bibinfo{person}{Djordje Jevdjic}, \bibinfo{person}{Junbo Deng}, \bibinfo{person}{Xingkun Yang}, \bibinfo{person}{Zhou Yu}, {and} \bibinfo{person}{Pengfei Zuo}.} \bibinfo{year}{2024}\natexlab{}.
\newblock \bibinfo{title}{Cost-Efficient Large Language Model Serving for Multi-turn Conversations with CachedAttention}.
\newblock
\showeprint[arxiv]{2403.19708}~[cs.CL]
\urldef\tempurl%
\url{https://arxiv.org/abs/2403.19708}
\showURL{%
\tempurl}


\bibitem[Gao et~al\mbox{.}(2025)]%
        {duetserve}
\bibfield{author}{\bibinfo{person}{Lei Gao}, \bibinfo{person}{Chaoyi Jiang}, \bibinfo{person}{Hossein~Entezari Zarch}, \bibinfo{person}{Daniel Wong}, {and} \bibinfo{person}{Murali Annavaram}.} \bibinfo{year}{2025}\natexlab{}.
\newblock \showarticletitle{DuetServe: Harmonizing Prefill and Decode for {LLM} Serving via Adaptive {GPU} Multiplexing}.
\newblock \bibinfo{journal}{\emph{arXiv preprint arXiv:2511.04791}} (\bibinfo{year}{2025}).
\newblock


\bibitem[Griggs et~al\mbox{.}(2024)]%
        {Melange}
\bibfield{author}{\bibinfo{person}{Tyler Griggs}, \bibinfo{person}{Xiaoxuan Liu}, \bibinfo{person}{Jiaxiang Yu}, \bibinfo{person}{Doyoung Kim}, \bibinfo{person}{Wei-Lin Chiang}, \bibinfo{person}{Alvin Cheung}, {and} \bibinfo{person}{Ion Stoica}.} \bibinfo{year}{2024}\natexlab{}.
\newblock \showarticletitle{M$\backslash$'elange: Cost efficient large language model serving by exploiting gpu heterogeneity}.
\newblock \bibinfo{journal}{\emph{arXiv preprint arXiv:2404.14527}} (\bibinfo{year}{2024}).
\newblock


\bibitem[He et~al\mbox{.}(2026)]%
        {ampd}
\bibfield{author}{\bibinfo{person}{Wenhao He}, \bibinfo{person}{Youhe Jiang}, \bibinfo{person}{Penghao Zhao}, \bibinfo{person}{Quanqing Xu}, \bibinfo{person}{Eiko Yoneki}, \bibinfo{person}{Bin Cui}, {and} \bibinfo{person}{Fangcheng Fu}.} \bibinfo{year}{2026}\natexlab{}.
\newblock \showarticletitle{Efficient multi-round llm inference over disaggregated serving}.
\newblock \bibinfo{journal}{\emph{arXiv preprint arXiv:2602.14516}} (\bibinfo{year}{2026}).
\newblock


\bibitem[Hu et~al\mbox{.}(2024)]%
        {memserve}
\bibfield{author}{\bibinfo{person}{Cunchen Hu}, \bibinfo{person}{Heyang Huang}, \bibinfo{person}{Junhao Hu}, \bibinfo{person}{Jiang Xu}, \bibinfo{person}{Xusheng Chen}, \bibinfo{person}{Tao Xie}, \bibinfo{person}{Chenxi Wang}, \bibinfo{person}{Sa Wang}, \bibinfo{person}{Yungang Bao}, \bibinfo{person}{Ninghui Sun}, {and} \bibinfo{person}{Yizhou Shan}.} \bibinfo{year}{2024}\natexlab{}.
\newblock \bibinfo{title}{MemServe: Context Caching for Disaggregated LLM Serving with Elastic Memory Pool}.
\newblock
\showeprint[arxiv]{2406.17565}~[cs.DC]
\urldef\tempurl%
\url{https://arxiv.org/abs/2406.17565}
\showURL{%
\tempurl}


\bibitem[Jain et~al\mbox{.}(2025)]%
        {jain-router}
\bibfield{author}{\bibinfo{person}{Kunal Jain}, \bibinfo{person}{Anjaly Parayil}, \bibinfo{person}{Ankur Mallick}, \bibinfo{person}{Esha Choukse}, \bibinfo{person}{Xiaoting Qin}, \bibinfo{person}{Jue Zhang}, \bibinfo{person}{Íñigo Goiri}, \bibinfo{person}{Rujia Wang}, \bibinfo{person}{Chetan Bansal}, \bibinfo{person}{Victor Rühle}, \bibinfo{person}{Anoop Kulkarni}, \bibinfo{person}{Steve Kofsky}, {and} \bibinfo{person}{Saravan Rajmohan}.} \bibinfo{year}{2025}\natexlab{}.
\newblock \bibinfo{title}{Intelligent Router for LLM Workloads: Improving Performance Through Workload-Aware Load Balancing}.
\newblock
\showeprint[arxiv]{2408.13510}~[cs.DC]
\urldef\tempurl%
\url{https://arxiv.org/abs/2408.13510}
\showURL{%
\tempurl}


\bibitem[Jiang et~al\mbox{.}(2025a)]%
        {ThunderServe}
\bibfield{author}{\bibinfo{person}{Youhe Jiang}, \bibinfo{person}{Fangcheng Fu}, \bibinfo{person}{Xiaozhe Yao}, \bibinfo{person}{Taiyi Wang}, \bibinfo{person}{Bin Cui}, \bibinfo{person}{Ana Klimovic}, {and} \bibinfo{person}{Eiko Yoneki}.} \bibinfo{year}{2025}\natexlab{a}.
\newblock \showarticletitle{ThunderServe: High-performance and Cost-efficient {LLM} Serving in Cloud Environments}. In \bibinfo{booktitle}{\emph{Proceedings of the Eighth Conference on Machine Learning and Systems, MLSys 2025, Santa Clara, CA, USA, May 12-15, 2025}}, \bibfield{editor}{\bibinfo{person}{Matei Zaharia}, \bibinfo{person}{Gauri Joshi}, {and} \bibinfo{person}{Yingyan~(Celine) Lin}} (Eds.). \bibinfo{publisher}{OpenReview.net/mlsys.org}.
\newblock


\bibitem[Jiang et~al\mbox{.}(2024)]%
        {HexGen}
\bibfield{author}{\bibinfo{person}{Youhe Jiang}, \bibinfo{person}{Ran Yan}, \bibinfo{person}{Xiaozhe Yao}, \bibinfo{person}{Yang Zhou}, \bibinfo{person}{Beidi Chen}, {and} \bibinfo{person}{Binhang Yuan}.} \bibinfo{year}{2024}\natexlab{}.
\newblock \showarticletitle{HexGen: Generative Inference of Large Language Model over Heterogeneous Environment}. In \bibinfo{booktitle}{\emph{Forty-first International Conference on Machine Learning, {ICML} 2024, Vienna, Austria, July 21-27, 2024}} \emph{(\bibinfo{series}{Proceedings of Machine Learning Research})}, \bibfield{editor}{\bibinfo{person}{Ruslan Salakhutdinov}, \bibinfo{person}{Zico Kolter}, \bibinfo{person}{Katherine~A. Heller}, \bibinfo{person}{Adrian Weller}, \bibinfo{person}{Nuria Oliver}, \bibinfo{person}{Jonathan Scarlett}, {and} \bibinfo{person}{Felix Berkenkamp}} (Eds.). \bibinfo{publisher}{{PMLR} / OpenReview.net}, \bibinfo{pages}{21946--21961}.
\newblock


\bibitem[Jiang et~al\mbox{.}(2025b)]%
        {HexGen-2}
\bibfield{author}{\bibinfo{person}{Youhe Jiang}, \bibinfo{person}{Ran Yan}, {and} \bibinfo{person}{Binhang Yuan}.} \bibinfo{year}{2025}\natexlab{b}.
\newblock \showarticletitle{HexGen-2: Disaggregated Generative Inference of LLMs in Heterogeneous Environment}. In \bibinfo{booktitle}{\emph{The Thirteenth International Conference on Learning Representations, {ICLR} 2025, Singapore, April 24-28, 2025}}. \bibinfo{publisher}{OpenReview.net}.
\newblock


\bibitem[Jimenez et~al\mbox{.}(2024)]%
        {swebench}
\bibfield{author}{\bibinfo{person}{Carlos~E. Jimenez}, \bibinfo{person}{John Yang}, \bibinfo{person}{Alexander Wettig}, \bibinfo{person}{Shunyu Yao}, \bibinfo{person}{Kexin Pei}, \bibinfo{person}{Ofir Press}, {and} \bibinfo{person}{Karthik Narasimhan}.} \bibinfo{year}{2024}\natexlab{}.
\newblock \bibinfo{title}{SWE-bench: Can Language Models Resolve Real-World GitHub Issues?}
\newblock
\showeprint[arxiv]{2310.06770}~[cs.CL]
\urldef\tempurl%
\url{https://arxiv.org/abs/2310.06770}
\showURL{%
\tempurl}


\bibitem[Kwon et~al\mbox{.}(2023)]%
        {vLLM}
\bibfield{author}{\bibinfo{person}{Woosuk Kwon}, \bibinfo{person}{Zhuohan Li}, \bibinfo{person}{Siyuan Zhuang}, \bibinfo{person}{Ying Sheng}, \bibinfo{person}{Lianmin Zheng}, \bibinfo{person}{Cody~Hao Yu}, \bibinfo{person}{Joseph Gonzalez}, \bibinfo{person}{Hao Zhang}, {and} \bibinfo{person}{Ion Stoica}.} \bibinfo{year}{2023}\natexlab{}.
\newblock \showarticletitle{Efficient Memory Management for Large Language Model Serving with PagedAttention}. In \bibinfo{booktitle}{\emph{Proceedings of the 29th Symposium on Operating Systems Principles, {SOSP} 2023, Koblenz, Germany, October 23-26, 2023}}, \bibfield{editor}{\bibinfo{person}{Jason Flinn}, \bibinfo{person}{Margo~I. Seltzer}, \bibinfo{person}{Peter Druschel}, \bibinfo{person}{Antoine Kaufmann}, {and} \bibinfo{person}{Jonathan Mace}} (Eds.). \bibinfo{publisher}{{ACM}}, \bibinfo{pages}{611--626}.
\newblock


\bibitem[Lai et~al\mbox{.}(2025)]%
        {tokenscale}
\bibfield{author}{\bibinfo{person}{Ruiqi Lai}, \bibinfo{person}{Hongrui Liu}, \bibinfo{person}{Chengzhi Lu}, \bibinfo{person}{Zonghao Liu}, \bibinfo{person}{Siyu Cao}, \bibinfo{person}{Siyang Shao}, \bibinfo{person}{Yixin Zhang}, \bibinfo{person}{Luo Mai}, {and} \bibinfo{person}{Dmitrii Ustiugov}.} \bibinfo{year}{2025}\natexlab{}.
\newblock \showarticletitle{TokenScale: Timely and Accurate Autoscaling for Disaggregated {LLM} Serving with Token Velocity}.
\newblock \bibinfo{journal}{\emph{arXiv preprint arXiv:2512.03416}} (\bibinfo{year}{2025}).
\newblock


\bibitem[Li et~al\mbox{.}(2025)]%
        {continuum}
\bibfield{author}{\bibinfo{person}{Hanchen Li}, \bibinfo{person}{Runyuan He}, \bibinfo{person}{Qiuyang Mang}, \bibinfo{person}{Qizheng Zhang}, \bibinfo{person}{Huanzhi Mao}, \bibinfo{person}{Xiaokun Chen}, \bibinfo{person}{Hangrui Zhou}, \bibinfo{person}{Alvin Cheung}, \bibinfo{person}{Joseph Gonzalez}, {and} \bibinfo{person}{Ion Stoica}.} \bibinfo{year}{2025}\natexlab{}.
\newblock \showarticletitle{Continuum: Efficient and robust multi-turn llm agent scheduling with kv cache time-to-live}.
\newblock \bibinfo{journal}{\emph{arXiv preprint arXiv:2511.02230}} (\bibinfo{year}{2025}).
\newblock


\bibitem[Li et~al\mbox{.}(2024)]%
        {snapkv}
\bibfield{author}{\bibinfo{person}{Yuhong Li}, \bibinfo{person}{Yingbing Huang}, \bibinfo{person}{Bowen Yang}, \bibinfo{person}{Bharat Venkitesh}, \bibinfo{person}{Acyr Locatelli}, \bibinfo{person}{Hanchen Ye}, \bibinfo{person}{Tianle Cai}, \bibinfo{person}{Patrick Lewis}, {and} \bibinfo{person}{Deming Chen}.} \bibinfo{year}{2024}\natexlab{}.
\newblock \bibinfo{title}{SnapKV: LLM Knows What You are Looking for Before Generation}.
\newblock
\showeprint[arxiv]{2404.14469}~[cs.CL]
\urldef\tempurl%
\url{https://arxiv.org/abs/2404.14469}
\showURL{%
\tempurl}


\bibitem[Li et~al\mbox{.}(2026)]%
        {ppd}
\bibfield{author}{\bibinfo{person}{Zongze Li}, \bibinfo{person}{Jingyu Liu}, \bibinfo{person}{Zhen Xu}, \bibinfo{person}{Yineng Zhang}, \bibinfo{person}{Tahseen Rabbani}, {and} \bibinfo{person}{Ce Zhang}.} \bibinfo{year}{2026}\natexlab{}.
\newblock \bibinfo{title}{Not All Prefills Are Equal: PPD Disaggregation for Multi-turn LLM Serving}.
\newblock
\showeprint[arxiv]{2603.13358}~[cs.NI]
\urldef\tempurl%
\url{https://arxiv.org/abs/2603.13358}
\showURL{%
\tempurl}


\bibitem[Lin et~al\mbox{.}(2024)]%
        {parrot}
\bibfield{author}{\bibinfo{person}{Chaofan Lin}, \bibinfo{person}{Zhenhua Han}, \bibinfo{person}{Chengruidong Zhang}, \bibinfo{person}{Yuqing Yang}, \bibinfo{person}{Fan Yang}, \bibinfo{person}{Chen Chen}, {and} \bibinfo{person}{Lili Qiu}.} \bibinfo{year}{2024}\natexlab{}.
\newblock \showarticletitle{Parrot: Efficient Serving of LLM-based Applications with Semantic Variable}. In \bibinfo{booktitle}{\emph{18th {USENIX} Symposium on Operating Systems Design and Implementation, {OSDI} 2024, Santa Clara, CA, USA, July 10-12, 2024}}, \bibfield{editor}{\bibinfo{person}{Ada Gavrilovska} {and} \bibinfo{person}{Douglas~B. Terry}} (Eds.). \bibinfo{publisher}{{USENIX} Association}, \bibinfo{pages}{929--945}.
\newblock


\bibitem[Liu et~al\mbox{.}(2024b)]%
        {AgentBench}
\bibfield{author}{\bibinfo{person}{Xiao Liu}, \bibinfo{person}{Hao Yu}, \bibinfo{person}{Hanchen Zhang}, \bibinfo{person}{Yifan Xu}, \bibinfo{person}{Xuanyu Lei}, \bibinfo{person}{Hanyu Lai}, \bibinfo{person}{Yu Gu}, \bibinfo{person}{Hangliang Ding}, \bibinfo{person}{Kaiwen Men}, \bibinfo{person}{Kejuan Yang}, \bibinfo{person}{Shudan Zhang}, \bibinfo{person}{Xiang Deng}, \bibinfo{person}{Aohan Zeng}, \bibinfo{person}{Zhengxiao Du}, \bibinfo{person}{Chenhui Zhang}, \bibinfo{person}{Sheng Shen}, \bibinfo{person}{Tianjun Zhang}, \bibinfo{person}{Yu Su}, \bibinfo{person}{Huan Sun}, \bibinfo{person}{Minlie Huang}, \bibinfo{person}{Yuxiao Dong}, {and} \bibinfo{person}{Jie Tang}.} \bibinfo{year}{2024}\natexlab{b}.
\newblock \showarticletitle{AgentBench: Evaluating LLMs as Agents}. In \bibinfo{booktitle}{\emph{The Twelfth International Conference on Learning Representations, {ICLR} 2024, Vienna, Austria, May 7-11, 2024}}. \bibinfo{publisher}{OpenReview.net}.
\newblock


\bibitem[Liu et~al\mbox{.}(2024a)]%
        {cachegen}
\bibfield{author}{\bibinfo{person}{Yuhan Liu}, \bibinfo{person}{Hanchen Li}, \bibinfo{person}{Yihua Cheng}, \bibinfo{person}{Siddhant Ray}, \bibinfo{person}{Yuyang Huang}, \bibinfo{person}{Qizheng Zhang}, \bibinfo{person}{Kuntai Du}, \bibinfo{person}{Jiayi Yao}, \bibinfo{person}{Shan Lu}, \bibinfo{person}{Ganesh Ananthanarayanan}, \bibinfo{person}{Michael Maire}, \bibinfo{person}{Henry Hoffmann}, \bibinfo{person}{Ari Holtzman}, {and} \bibinfo{person}{Junchen Jiang}.} \bibinfo{year}{2024}\natexlab{a}.
\newblock \showarticletitle{CacheGen: {KV} Cache Compression and Streaming for Fast Large Language Model Serving}. In \bibinfo{booktitle}{\emph{Proceedings of the {ACM} {SIGCOMM} 2024 Conference, {ACM} {SIGCOMM} 2024, Sydney, NSW, Australia, August 4-8, 2024}}. \bibinfo{publisher}{{ACM}}, \bibinfo{pages}{38--56}.
\newblock


\bibitem[Luo et~al\mbox{.}(2025)]%
        {autellix}
\bibfield{author}{\bibinfo{person}{Michael Luo}, \bibinfo{person}{Xiaoxiang Shi}, \bibinfo{person}{Colin Cai}, \bibinfo{person}{Tianjun Zhang}, \bibinfo{person}{Justin Wong}, \bibinfo{person}{Yichuan Wang}, \bibinfo{person}{Chi Wang}, \bibinfo{person}{Yanping Huang}, \bibinfo{person}{Zhifeng Chen}, \bibinfo{person}{Joseph~E. Gonzalez}, {and} \bibinfo{person}{Ion Stoica}.} \bibinfo{year}{2025}\natexlab{}.
\newblock \bibinfo{title}{Autellix: An Efficient Serving Engine for LLM Agents as General Programs}.
\newblock
\showeprint[arxiv]{2502.13965}~[cs.LG]
\urldef\tempurl%
\url{https://arxiv.org/abs/2502.13965}
\showURL{%
\tempurl}


\bibitem[Mei et~al\mbox{.}(2025)]%
        {Helix}
\bibfield{author}{\bibinfo{person}{Yixuan Mei}, \bibinfo{person}{Yonghao Zhuang}, \bibinfo{person}{Xupeng Miao}, \bibinfo{person}{Juncheng Yang}, \bibinfo{person}{Zhihao Jia}, {and} \bibinfo{person}{Rashmi Vinayak}.} \bibinfo{year}{2025}\natexlab{}.
\newblock \showarticletitle{Helix: Serving Large Language Models over Heterogeneous GPUs and Network via Max-Flow}. In \bibinfo{booktitle}{\emph{Proceedings of the 30th {ACM} International Conference on Architectural Support for Programming Languages and Operating Systems, Volume 1, {ASPLOS} 2025, Rotterdam, The Netherlands, 30 March 2025 - 3 April 2025}}, \bibfield{editor}{\bibinfo{person}{Lieven Eeckhout}, \bibinfo{person}{Georgios Smaragdakis}, \bibinfo{person}{Kaitai Liang}, \bibinfo{person}{Adrian Sampson}, \bibinfo{person}{Martha~A. Kim}, {and} \bibinfo{person}{Christopher~J. Rossbach}} (Eds.). \bibinfo{publisher}{{ACM}}, \bibinfo{pages}{586--602}.
\newblock


\bibitem[{NVIDIA}(2025)]%
        {dynamo}
\bibfield{author}{\bibinfo{person}{{NVIDIA}}.} \bibinfo{year}{2025}\natexlab{}.
\newblock \bibinfo{title}{{NVIDIA} {Dynamo}: A Low-Latency Distributed Inference Framework for Scaling Reasoning {AI} Models}.
\newblock \bibinfo{howpublished}{\url{https://developer.nvidia.com/dynamo}}.
\newblock


\bibitem[Pan et~al\mbox{.}(2025)]%
        {kvflow}
\bibfield{author}{\bibinfo{person}{Zaifeng Pan}, \bibinfo{person}{Ajjkumar Patel}, \bibinfo{person}{Zhengding Hu}, \bibinfo{person}{Yipeng Shen}, \bibinfo{person}{Yue Guan}, \bibinfo{person}{Wan-Lu Li}, \bibinfo{person}{Lianhui Qin}, \bibinfo{person}{Yida Wang}, {and} \bibinfo{person}{Yufei Ding}.} \bibinfo{year}{2025}\natexlab{}.
\newblock \bibinfo{title}{KVFlow: Efficient Prefix Caching for Accelerating LLM-Based Multi-Agent Workflows}.
\newblock
\showeprint[arxiv]{2507.07400}~[cs.DC]
\urldef\tempurl%
\url{https://arxiv.org/abs/2507.07400}
\showURL{%
\tempurl}


\bibitem[Patel et~al\mbox{.}(2024)]%
        {Splitwise}
\bibfield{author}{\bibinfo{person}{Pratyush Patel}, \bibinfo{person}{Esha Choukse}, \bibinfo{person}{Chaojie Zhang}, \bibinfo{person}{Aashaka Shah}, \bibinfo{person}{{\'{I}}{\~{n}}igo Goiri}, \bibinfo{person}{Saeed Maleki}, {and} \bibinfo{person}{Ricardo Bianchini}.} \bibinfo{year}{2024}\natexlab{}.
\newblock \showarticletitle{Splitwise: Efficient Generative {LLM} Inference Using Phase Splitting}. In \bibinfo{booktitle}{\emph{51st {ACM/IEEE} Annual International Symposium on Computer Architecture, {ISCA} 2024, Buenos Aires, Argentina, June 29 - July 3, 2024}}. \bibinfo{publisher}{{IEEE}}, \bibinfo{pages}{118--132}.
\newblock


\bibitem[Pope et~al\mbox{.}(2023)]%
        {PopeDCDBHXAD23}
\bibfield{author}{\bibinfo{person}{Reiner Pope}, \bibinfo{person}{Sholto Douglas}, \bibinfo{person}{Aakanksha Chowdhery}, \bibinfo{person}{Jacob Devlin}, \bibinfo{person}{James Bradbury}, \bibinfo{person}{Jonathan Heek}, \bibinfo{person}{Kefan Xiao}, \bibinfo{person}{Shivani Agrawal}, {and} \bibinfo{person}{Jeff Dean}.} \bibinfo{year}{2023}\natexlab{}.
\newblock \showarticletitle{Efficiently Scaling Transformer Inference}. In \bibinfo{booktitle}{\emph{Proceedings of the Sixth Conference on Machine Learning and Systems, MLSys 2023, Miami, FL, USA, June 4-8, 2023}}, \bibfield{editor}{\bibinfo{person}{Dawn Song}, \bibinfo{person}{Michael Carbin}, {and} \bibinfo{person}{Tianqi Chen}} (Eds.). \bibinfo{publisher}{mlsys.org}.
\newblock


\bibitem[Qin et~al\mbox{.}(2025)]%
        {mooncake}
\bibfield{author}{\bibinfo{person}{Ruoyu Qin}, \bibinfo{person}{Zheming Li}, \bibinfo{person}{Weiran He}, \bibinfo{person}{Jialei Cui}, \bibinfo{person}{Feng Ren}, \bibinfo{person}{Mingxing Zhang}, \bibinfo{person}{Yongwei Wu}, \bibinfo{person}{Weimin Zheng}, {and} \bibinfo{person}{Xinran Xu}.} \bibinfo{year}{2025}\natexlab{}.
\newblock \showarticletitle{Mooncake: Trading More Storage for Less Computation - {A} KVCache-centric Architecture for Serving {LLM} Chatbot}. In \bibinfo{booktitle}{\emph{23rd {USENIX} Conference on File and Storage Technologies, {FAST} 2025, Santa Clara, CA, February 25-27, 2025}}, \bibfield{editor}{\bibinfo{person}{Haryadi~S. Gunawi} {and} \bibinfo{person}{Vasily Tarasov}} (Eds.). \bibinfo{publisher}{{USENIX} Association}, \bibinfo{pages}{155--170}.
\newblock


\bibitem[Qin et~al\mbox{.}(2024)]%
        {ToolLLM}
\bibfield{author}{\bibinfo{person}{Yujia Qin}, \bibinfo{person}{Shihao Liang}, \bibinfo{person}{Yining Ye}, \bibinfo{person}{Kunlun Zhu}, \bibinfo{person}{Lan Yan}, \bibinfo{person}{Yaxi Lu}, \bibinfo{person}{Yankai Lin}, \bibinfo{person}{Xin Cong}, \bibinfo{person}{Xiangru Tang}, \bibinfo{person}{Bill Qian}, \bibinfo{person}{Sihan Zhao}, \bibinfo{person}{Lauren Hong}, \bibinfo{person}{Runchu Tian}, \bibinfo{person}{Ruobing Xie}, \bibinfo{person}{Jie Zhou}, \bibinfo{person}{Mark Gerstein}, \bibinfo{person}{Dahai Li}, \bibinfo{person}{Zhiyuan Liu}, {and} \bibinfo{person}{Maosong Sun}.} \bibinfo{year}{2024}\natexlab{}.
\newblock \showarticletitle{ToolLLM: Facilitating Large Language Models to Master 16000+ Real-world APIs}. In \bibinfo{booktitle}{\emph{The Twelfth International Conference on Learning Representations, {ICLR} 2024, Vienna, Austria, May 7-11, 2024}}. \bibinfo{publisher}{OpenReview.net}.
\newblock


\bibitem[Radford et~al\mbox{.}(2018)]%
        {gpt-1}
\bibfield{author}{\bibinfo{person}{Alec Radford}, \bibinfo{person}{Karthik Narasimhan}, \bibinfo{person}{Tim Salimans}, \bibinfo{person}{Ilya Sutskever}, {et~al\mbox{.}}} \bibinfo{year}{2018}\natexlab{}.
\newblock \showarticletitle{Improving language understanding by generative pre-training}.
\newblock  (\bibinfo{year}{2018}).
\newblock
\urldef\tempurl%
\url{https://cdn.openai.com/research-covers/language-unsupervised/language_understanding_paper.pdf}
\showURL{%
\tempurl}


\bibitem[Radford et~al\mbox{.}(2019)]%
        {gpt-2}
\bibfield{author}{\bibinfo{person}{Alec Radford}, \bibinfo{person}{Jeffrey Wu}, \bibinfo{person}{Rewon Child}, \bibinfo{person}{David Luan}, \bibinfo{person}{Dario Amodei}, \bibinfo{person}{Ilya Sutskever}, {et~al\mbox{.}}} \bibinfo{year}{2019}\natexlab{}.
\newblock \showarticletitle{Language models are unsupervised multitask learners}.
\newblock  (\bibinfo{year}{2019}).
\newblock
\urldef\tempurl%
\url{https://cdn.openai.com/better-language-models/language_models_are_unsupervised_multitask_learners.pdf}
\showURL{%
\tempurl}


\bibitem[Richards(2023)]%
        {AutoGPT}
\bibfield{author}{\bibinfo{person}{Toran~Bruce Richards}.} \bibinfo{year}{2023}\natexlab{}.
\newblock \bibinfo{booktitle}{\emph{AutoGPT}}.
\newblock
\urldef\tempurl%
\url{https://github.com/Significant-Gravitas/AutoGPT}
\showURL{%
\tempurl}


\bibitem[Schick et~al\mbox{.}(2023)]%
        {Toolformer}
\bibfield{author}{\bibinfo{person}{Timo Schick}, \bibinfo{person}{Jane Dwivedi{-}Yu}, \bibinfo{person}{Roberto Dess{\`{\i}}}, \bibinfo{person}{Roberta Raileanu}, \bibinfo{person}{Maria Lomeli}, \bibinfo{person}{Eric Hambro}, \bibinfo{person}{Luke Zettlemoyer}, \bibinfo{person}{Nicola Cancedda}, {and} \bibinfo{person}{Thomas Scialom}.} \bibinfo{year}{2023}\natexlab{}.
\newblock \showarticletitle{Toolformer: Language Models Can Teach Themselves to Use Tools}. In \bibinfo{booktitle}{\emph{Advances in Neural Information Processing Systems 36: Annual Conference on Neural Information Processing Systems 2023, NeurIPS 2023, New Orleans, LA, USA, December 10 - 16, 2023}}, \bibfield{editor}{\bibinfo{person}{Alice Oh}, \bibinfo{person}{Tristan Naumann}, \bibinfo{person}{Amir Globerson}, \bibinfo{person}{Kate Saenko}, \bibinfo{person}{Moritz Hardt}, {and} \bibinfo{person}{Sergey Levine}} (Eds.).
\newblock


\bibitem[Shi et~al\mbox{.}(2024)]%
        {greenllm}
\bibfield{author}{\bibinfo{person}{Tianyao Shi}, \bibinfo{person}{Yanran Wu}, \bibinfo{person}{Sihang Liu}, {and} \bibinfo{person}{Yi Ding}.} \bibinfo{year}{2024}\natexlab{}.
\newblock \bibinfo{title}{GreenLLM: Disaggregating Large Language Model Serving on Heterogeneous GPUs for Lower Carbon Emissions}.
\newblock
\showeprint[arxiv]{2412.20322}~[cs.AR]
\urldef\tempurl%
\url{https://arxiv.org/abs/2412.20322}
\showURL{%
\tempurl}


\bibitem[Shi et~al\mbox{.}(2025)]%
        {nexus}
\bibfield{author}{\bibinfo{person}{Xiaoxiang Shi}, \bibinfo{person}{Colin Cai}, \bibinfo{person}{Junjia Du}, {and} \bibinfo{person}{Zhihao Jia}.} \bibinfo{year}{2025}\natexlab{}.
\newblock \showarticletitle{Nexus: Proactive Intra-{GPU} Disaggregation of Prefill and Decode in {LLM} Serving}.
\newblock \bibinfo{journal}{\emph{arXiv preprint arXiv:2507.06608}} (\bibinfo{year}{2025}).
\newblock


\bibitem[Shinn et~al\mbox{.}(2023)]%
        {Reflexion}
\bibfield{author}{\bibinfo{person}{Noah Shinn}, \bibinfo{person}{Federico Cassano}, \bibinfo{person}{Ashwin Gopinath}, \bibinfo{person}{Karthik Narasimhan}, {and} \bibinfo{person}{Shunyu Yao}.} \bibinfo{year}{2023}\natexlab{}.
\newblock \showarticletitle{Reflexion: language agents with verbal reinforcement learning}. In \bibinfo{booktitle}{\emph{Advances in Neural Information Processing Systems 36: Annual Conference on Neural Information Processing Systems 2023, NeurIPS 2023, New Orleans, LA, USA, December 10 - 16, 2023}}, \bibfield{editor}{\bibinfo{person}{Alice Oh}, \bibinfo{person}{Tristan Naumann}, \bibinfo{person}{Amir Globerson}, \bibinfo{person}{Kate Saenko}, \bibinfo{person}{Moritz Hardt}, {and} \bibinfo{person}{Sergey Levine}} (Eds.).
\newblock


\bibitem[Stojkovic et~al\mbox{.}(2025)]%
        {dynamoLLM}
\bibfield{author}{\bibinfo{person}{Jovan Stojkovic}, \bibinfo{person}{Chaojie Zhang}, \bibinfo{person}{Íñigo Goiri}, \bibinfo{person}{Josep Torrellas}, {and} \bibinfo{person}{Esha Choukse}.} \bibinfo{year}{2025}\natexlab{}.
\newblock \showarticletitle{DynamoLLM: Designing LLM Inference Clusters for Performance and Energy Efficiency}. In \bibinfo{booktitle}{\emph{2025 IEEE International Symposium on High Performance Computer Architecture (HPCA)}}. \bibinfo{publisher}{IEEE}, \bibinfo{pages}{1348–1362}.
\newblock
\href{https://doi.org/10.1109/hpca61900.2025.00102}{doi:\nolinkurl{10.1109/hpca61900.2025.00102}}


\bibitem[Sun et~al\mbox{.}(2024)]%
        {llumnix}
\bibfield{author}{\bibinfo{person}{Biao Sun}, \bibinfo{person}{Ziming Huang}, \bibinfo{person}{Hanyu Zhao}, \bibinfo{person}{Wencong Xiao}, \bibinfo{person}{Xinyi Zhang}, \bibinfo{person}{Yong Li}, {and} \bibinfo{person}{Wei Lin}.} \bibinfo{year}{2024}\natexlab{}.
\newblock \showarticletitle{Llumnix: Dynamic Scheduling for Large Language Model Serving}. In \bibinfo{booktitle}{\emph{18th USENIX Symposium on Operating Systems Design and Implementation (OSDI 24)}}. \bibinfo{publisher}{USENIX Association}, \bibinfo{pages}{173--191}.
\newblock


\bibitem[Vaswani et~al\mbox{.}(2017)]%
        {transformer}
\bibfield{author}{\bibinfo{person}{Ashish Vaswani}, \bibinfo{person}{Noam Shazeer}, \bibinfo{person}{Niki Parmar}, \bibinfo{person}{Jakob Uszkoreit}, \bibinfo{person}{Llion Jones}, \bibinfo{person}{Aidan~N. Gomez}, \bibinfo{person}{Lukasz Kaiser}, {and} \bibinfo{person}{Illia Polosukhin}.} \bibinfo{year}{2017}\natexlab{}.
\newblock \showarticletitle{Attention is All you Need}. In \bibinfo{booktitle}{\emph{Advances in Neural Information Processing Systems 30: Annual Conference on Neural Information Processing Systems 2017, December 4-9, 2017, Long Beach, CA, {USA}}}, \bibfield{editor}{\bibinfo{person}{Isabelle Guyon}, \bibinfo{person}{Ulrike von Luxburg}, \bibinfo{person}{Samy Bengio}, \bibinfo{person}{Hanna~M. Wallach}, \bibinfo{person}{Rob Fergus}, \bibinfo{person}{S.~V.~N. Vishwanathan}, {and} \bibinfo{person}{Roman Garnett}} (Eds.). \bibinfo{pages}{5998--6008}.
\newblock


\bibitem[Wang et~al\mbox{.}(2025)]%
        {taichi}
\bibfield{author}{\bibinfo{person}{Chao Wang}, \bibinfo{person}{Pengfei Zuo}, \bibinfo{person}{Zhangyu Chen}, \bibinfo{person}{Yunkai Liang}, \bibinfo{person}{Zhou Yu}, {and} \bibinfo{person}{Ming-Chang Yang}.} \bibinfo{year}{2025}\natexlab{}.
\newblock \bibinfo{title}{Prefill-Decode Aggregation or Disaggregation? Unifying Both for Goodput-Optimized LLM Serving}.
\newblock
\showeprint[arxiv]{2508.01989}~[cs.DC]
\urldef\tempurl%
\url{https://arxiv.org/abs/2508.01989}
\showURL{%
\tempurl}


\bibitem[Yao et~al\mbox{.}(2023)]%
        {ReAct}
\bibfield{author}{\bibinfo{person}{Shunyu Yao}, \bibinfo{person}{Jeffrey Zhao}, \bibinfo{person}{Dian Yu}, \bibinfo{person}{Nan Du}, \bibinfo{person}{Izhak Shafran}, \bibinfo{person}{Karthik~R. Narasimhan}, {and} \bibinfo{person}{Yuan Cao}.} \bibinfo{year}{2023}\natexlab{}.
\newblock \showarticletitle{ReAct: Synergizing Reasoning and Acting in Language Models}. In \bibinfo{booktitle}{\emph{The Eleventh International Conference on Learning Representations, {ICLR} 2023, Kigali, Rwanda, May 1-5, 2023}}. \bibinfo{publisher}{OpenReview.net}.
\newblock


\bibitem[Yu et~al\mbox{.}(2022)]%
        {orca}
\bibfield{author}{\bibinfo{person}{Gyeong{-}In Yu}, \bibinfo{person}{Joo~Seong Jeong}, \bibinfo{person}{Geon{-}Woo Kim}, \bibinfo{person}{Soojeong Kim}, {and} \bibinfo{person}{Byung{-}Gon Chun}.} \bibinfo{year}{2022}\natexlab{}.
\newblock \showarticletitle{Orca: {A} Distributed Serving System for Transformer-Based Generative Models}. In \bibinfo{booktitle}{\emph{16th {USENIX} Symposium on Operating Systems Design and Implementation, {OSDI} 2022, Carlsbad, CA, USA, July 11-13, 2022}}, \bibfield{editor}{\bibinfo{person}{Marcos~K. Aguilera} {and} \bibinfo{person}{Hakim Weatherspoon}} (Eds.). \bibinfo{publisher}{{USENIX} Association}, \bibinfo{pages}{521--538}.
\newblock


\bibitem[Zhang et~al\mbox{.}(2025)]%
        {blitzscale}
\bibfield{author}{\bibinfo{person}{Dingyan Zhang}, \bibinfo{person}{Haotian Wang}, \bibinfo{person}{Yang Liu}, \bibinfo{person}{Xingda Wei}, \bibinfo{person}{Yizhou Shan}, \bibinfo{person}{Rong Chen}, {and} \bibinfo{person}{Haibo Chen}.} \bibinfo{year}{2025}\natexlab{}.
\newblock \showarticletitle{BlitzScale: Fast and Live Large Model Autoscaling with O(1) Host Caching}. In \bibinfo{booktitle}{\emph{19th USENIX Symposium on Operating Systems Design and Implementation (OSDI 25)}}. \bibinfo{publisher}{USENIX Association}, \bibinfo{pages}{275--293}.
\newblock


\bibitem[Zheng et~al\mbox{.}(2024)]%
        {SGLang}
\bibfield{author}{\bibinfo{person}{Lianmin Zheng}, \bibinfo{person}{Liangsheng Yin}, \bibinfo{person}{Zhiqiang Xie}, \bibinfo{person}{Chuyue Sun}, \bibinfo{person}{Jeff Huang}, \bibinfo{person}{Cody~Hao Yu}, \bibinfo{person}{Shiyi Cao}, \bibinfo{person}{Christos Kozyrakis}, \bibinfo{person}{Ion Stoica}, \bibinfo{person}{Joseph~E. Gonzalez}, \bibinfo{person}{Clark~W. Barrett}, {and} \bibinfo{person}{Ying Sheng}.} \bibinfo{year}{2024}\natexlab{}.
\newblock \showarticletitle{SGLang: Efficient Execution of Structured Language Model Programs}. In \bibinfo{booktitle}{\emph{Advances in Neural Information Processing Systems 38: Annual Conference on Neural Information Processing Systems 2024, NeurIPS 2024, Vancouver, BC, Canada, December 10 - 15, 2024}}, \bibfield{editor}{\bibinfo{person}{Amir Globersons}, \bibinfo{person}{Lester Mackey}, \bibinfo{person}{Danielle Belgrave}, \bibinfo{person}{Angela Fan}, \bibinfo{person}{Ulrich Paquet}, \bibinfo{person}{Jakub~M. Tomczak}, {and} \bibinfo{person}{Cheng Zhang}} (Eds.).
\newblock


\bibitem[Zhong et~al\mbox{.}(2024)]%
        {DistServe}
\bibfield{author}{\bibinfo{person}{Yinmin Zhong}, \bibinfo{person}{Shengyu Liu}, \bibinfo{person}{Junda Chen}, \bibinfo{person}{Jianbo Hu}, \bibinfo{person}{Yibo Zhu}, \bibinfo{person}{Xuanzhe Liu}, \bibinfo{person}{Xin Jin}, {and} \bibinfo{person}{Hao Zhang}.} \bibinfo{year}{2024}\natexlab{}.
\newblock \showarticletitle{DistServe: Disaggregating Prefill and Decoding for Goodput-optimized Large Language Model Serving}. In \bibinfo{booktitle}{\emph{18th {USENIX} Symposium on Operating Systems Design and Implementation, {OSDI} 2024, Santa Clara, CA, USA, July 10-12, 2024}}, \bibfield{editor}{\bibinfo{person}{Ada Gavrilovska} {and} \bibinfo{person}{Douglas~B. Terry}} (Eds.). \bibinfo{publisher}{{USENIX} Association}, \bibinfo{pages}{193--210}.
\newblock


\end{thebibliography}

\end{document}